\documentclass[
superscriptaddress,
preprint,
showpacs,preprintnumbers,
 amsmath, amssymb,
 aps, physrev,
]{revtex4-2}

\usepackage{graphicx}
\usepackage{dcolumn}
\usepackage{bm}

\begin{document}


\title{\textbf{Enhanced third-harmonic generation and degenerate four-wave mixing in an all-dielectric metasurfaces via Brillouin zone folding-induced bound states in the continuum} }%

\author{ Meibao Qin }
\email{qinmb@ncpu.edu.cn}
\affiliation{School of Education, Nanchang Institute of Science and Technology, Nanchang 330108, China.}

\author{ Feng Wu}
 \affiliation{School of Optoelectronic Engineering, Guangdong Polytechnic Normal University, Guangzhou 510665, China}%
 
 \author{Tingting Liu}
\affiliation{School of Information Engineering, Nanchang University, Nanchang 330031, China}

\affiliation{Institute for Advanced Study, Nanchang University, Nanchang 330031, China}
 
\author{ Dandan Zhang}
 \affiliation{School of Physics and Materials Science, Nanchang University, Nanchang 330031, China}%
 
\author{Shuyuan Xiao}
\email{syxiao@ncu.edu.cn}
\affiliation{School of Information Engineering, Nanchang University, Nanchang 330031, China}

\affiliation{Institute for Advanced Study, Nanchang University, Nanchang 330031, China}


\date{\today}

\begin{abstract}
Bound states in the continuum (BICs) exhibit significant electric field confinement capabilities and have recently been employed to enhance nonlinear optics response at the nanoscale. In this study, we achieve substantial enhancement of third-harmonic generation (THG) and degenerate four-wave mixing (dFWM) by implementing Brillouin zone folding-induced BICs (BZF-BICs) in an air-hole type nonlinear metasurface. By introducing gap perturbations within the metasurface, guided modes below the light line can be folded into the light cone, resulting in three resonant modes: guided resonances (GRs), $\Gamma$-BICs, and BZF-BICs. Through the eigenvalue analysis and multipole decompositions, we establish their excitation conditions. With their resonantly enhanced local field,  we successfully boost both THG and dFWM under $x$- and $y$- polarizations within the same metasurfaces. The simulated results indicate that the BZF-BICs provide the most significant enhancement of third-order nonlinear optical responses, with the output power of THG to 10$^{-4}$ W and dFWM output power of 10$^{-2}$ W under a moderate input power density of 1 MW/cm$^{2}$.  These findings demonstrate that the BZF-BICs can offer an effective pathway for chip-scale nonlinear optical applications.
 
\end{abstract}

\maketitle


\section{\label{sec:level1}INTRODUCTION}

Third-harmonic generation (THG) and degenerate four-wave mixing (dFWM) are both third-order nonlinear optical phenomena that arise from the third-order term in the power series expansion of a material polarization to an applied electric field \cite{Boyd2020,sheik2000third}. These processes have been utilized in a wide range of optical applications, including telecommunications, quantum computing, and imaging technologies \cite{dinu2003third, Hutchinson2004, Xu2019, gao2018nonlinear}. Early studies on third-order nonlinear optical effects primarily focused on specific crystal materials with birefringent properties, oscillating mismatches artificially “quasi-phase matching” crystal, and two-dimensional nonlinear photonic quasicrystals constructed using the generalized dual-grid method \cite{Maker1962, Lifshitz2005}. These approaches originate from the dispersion effects present in bulk materials, which causes the propagation velocities of the fundamental and octave frequencies within the crystal to be incongruent, resulting in a phenomenon known as phase mismatch, whereby the intensity of the octave signal fails to accumulate \cite{Boyd2020, Paul2003}. However, this effect can be negligible when the size of the optical micro-cavity is smaller than the wavelength of incident light \cite{Krasnok2018, Huang2020}. Therefore, optical metasurfaces with sub-wavelength dimensions are emerging as a novel avenue for exploring nonlinear optical effects and applications \cite{kivshar2018all, Liu2022, Vabishchevich2023, Xu2024, Wang2024}.

According to the classification of constituent materials,  optical metasurfaces are primarily categorized into metallic and dielectric types. While the metallic metasurfaces exhibit significant near-field enhancement via surface plasmons, ohmic losses in the visible range limit the development of practical optical devices \cite{Panoiu2018, BinAlam2021, Liu2023}. Dielectric metasurfaces support abundant Mie-type resonant modes, making them a lossless platforms for enhancing and modulating nonlinear phenomena \cite{Yang2015, Hu2020}. However, they are often hindered by low efficiency in nonlinear frequency conversion. Under the circumstances, bound states in the continuum (BICs) in all-dielectric metasurfaces present a promising strategy for enhancing nonlinear efficiency, owing to their exceptional capabilities for light-field confinement and controllable radiation loss \cite{Alam2018,  MelikGaykazyan2019, Koshelev2019, Qi2023, Sun2024, Tu2024}. BIC is a dark state that exists in a continuous domain and typically cannot be directly excited by incident light due to symmetry mismatch or interference phase cancellation \cite{hsu2016bound, friedrich1985interfering, koshelev2019nonradiating}. However, when the in-plane symmetry is broken or a plane wave is incident at an oblique angle, the BIC can be transformed into a quasi-BIC with a controllable quality ($Q$) factor, serving as an ideal platform for enhancing light-matter interactions in both linear and nonlinear regimes\cite{Wang2020,  Wu2021, Qin2021, Xie2021, Zeng2021, qu2022giant, liu2023boosting, huang2023resonant, Liu2024, Zhang2024, Liu2024a, hajian2024quasi, jiang2024tunable}.

For the exploration and optimization of optical structures, BIC nonlinear metasurfaces are increasingly employed to enhance the efficiency of the THG and dFWM processes \cite{Carletti2019, Ning2020, Xiao2022, Liu2023a, Tang2024, Qin2024}. However, as the wave vector $k_x$ increases in momentum space, the $Q$ factor of most BICs experiences a sharp decline, which limits their effectiveness in practical applications involving BIC-based optical devices \cite{shi2022terahertz, yan2024brillouin}. In sharp contrast, a novel kind of BIC, named Brillouin zone folding-induced BICs  (BZF-BICs), demonstrates a high $Q$ factor across a broader range in $k$ space and exhibits robustness against structural disorders \cite{Wang2023}. Recent studies have begun to investigate the use of BZF-BIC to enhance the light-matter interactions in the linear regime, with applications including highly coherent thermal emissions, near-perfect circular response, and enhanced intrinsic chirality \cite{sun2024exploiting, yan2024brillouin, Zhang2024}. Moreover, the robust $Q$ factor of BZF-BIC is particularly important for frequency conversion in nonlinear optics. However, research on nonlinear frequency conversion enhanced by BZF-BIC remains scarce.

In this work, we construct an air-hole-type nonlinear metasurfaces that integrates the concepts of GRs, $\Gamma$-BICs, and BZF-BICs to enhance and tailor the THG and dFWM within the structure. By analyzing the eigenmode in the first Brillouin zone, we elucidate the optical excitation condition required to excite these three modes, thereby providing a foundation for achieving third-order nonlinear optical phenomena. We then simulate the THG enhancement under the $x$ and $y$ polarization incident plane wave, finding that the maximum enhancement output power of the BZF-BIC reaches 10$^{-4}$ W when the input pumps are fixed at 1 MW/cm$^{2}$. Under identical input pump conditions, the output power of dFWM induced by BZF-BIC under $x$-polarization can be directly increased to 10$^{-2}$ W. This work suggests that BZF-BICs provide significant enhancement and effective modulation of the third-order nonlinear responses.

\section{\label{sec:level2}MODE ANALYSIS OF BICs}

\begin{figure}[htbp]
\centering
\includegraphics
[scale=0.6]{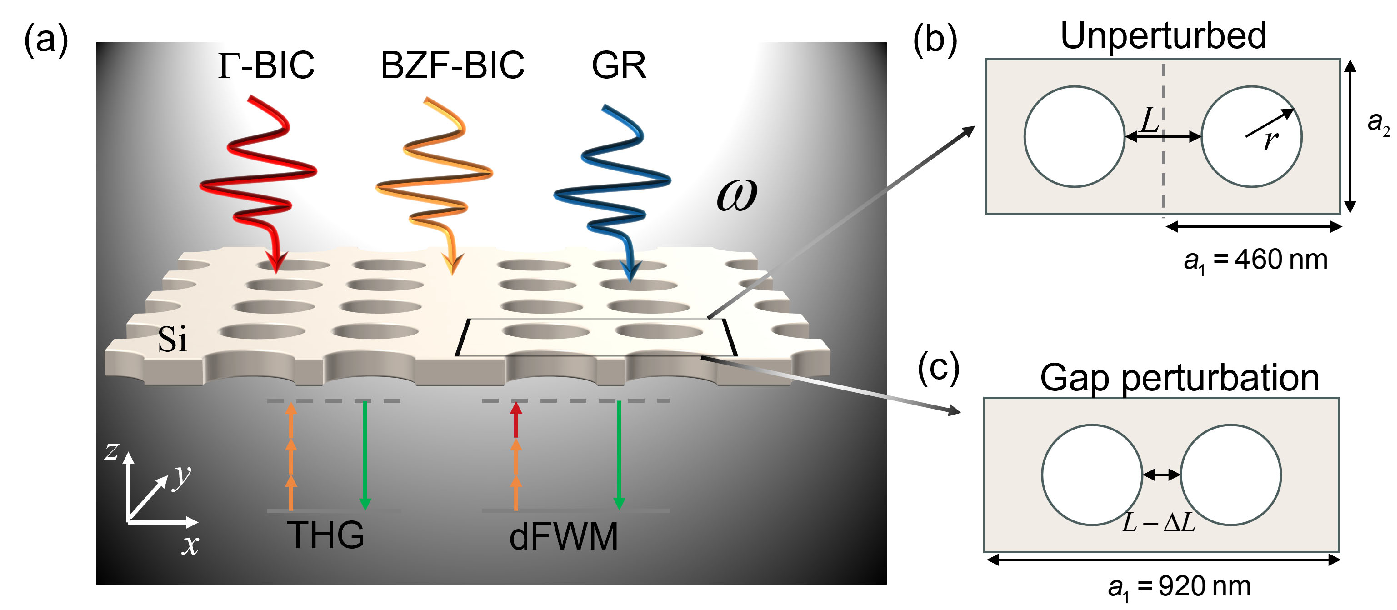}
\caption{\label{fig1}Enhanced third harmonic generation and degenerate four-wave mixing in all-dielectric nonlinear metasurfaces. (a) Schematic of a nonlinear metasurface composed of a Si membrane featuring a rectangular array of circular air holes, with two fixed geometric parameters: a height of $h=330$ nm, a radius of $r=130$ nm. (b) A unit cell of an unperturbed metasurface, characterized by periods of $a_{1}=460$ nm in the $x$ direction and $a_{2}=400$ nm in the $y$ direction, with a gap between the air hole in the $x$ direction of $L=200$ nm. (c) A unit cell of a gap-perturbed metasurface, with a period $a_{1}=920$ nm, and a variable asymmetry parameter defined as $\bm{\alpha}=\Delta L/L$.}
\end{figure}

We start from a nonlinear metasurface consisting of a silicon (Si) membrane that features a rectangular array of circular air holes. This structure is characterized by two fixed geometric parameters: a height of $h=330$ nm and a radius of $r=130$ nm, and it is suspended in air, as illustrated in Fig.~\ref{fig1}. The unperturbed unit cell is defined by periods of $a_{1}=460$ nm in the $x$ direction and $a_{2}=400$ nm in the $y$ direction, with a gap $L=200$ nm between the air hole in the  $x$ direction, as shown in Fig.~\ref{fig1} (a). To study the BZF-BIC, we introduce a gap perturbation $\Delta L$ in the $x$ direction, resulting in a modified gap of $L-\Delta L$ and changing $a_{1}$ to 920 nm, as depicted in Fig.~\ref{fig1} (b). To quantify the periodic perturbation, we define a variable asymmetry parameter as $\bm{\alpha}=\Delta L/L$. When the asymmetry parameter $\bm{\alpha}=0$, the unit cell of the nonlinear metasurface remains unperturbed. However, with the introduction of a gap perturbation $\bm{\alpha} \neq 0$, the unit cell size is doubled, and the first Brillouin zone is reduced to half of its original one. This change facilitates the transition of GMs from non-radiative dark mode to radiative resonances, resulting in the formation of GRs or BICs. Both the linear and nonlinear optical processes of the metasurface have been analyzed using 
the finite element method (FEM) implemented in the commercially available software of COMSOL Multiphysics. For the simulation, we adopt periodic boundary conditions in the $x$ and $y$ directions, while a perfect matching layer (PML) is set in the $z$ direction. The refractive index of Si is set to 3.5, with the imaginary part ignored in the eigenmode calculation, however, it will be considered in the THG and dFWM calculations that follow.

\begin{figure}[htbp]
\centering
\includegraphics
[scale=0.6]{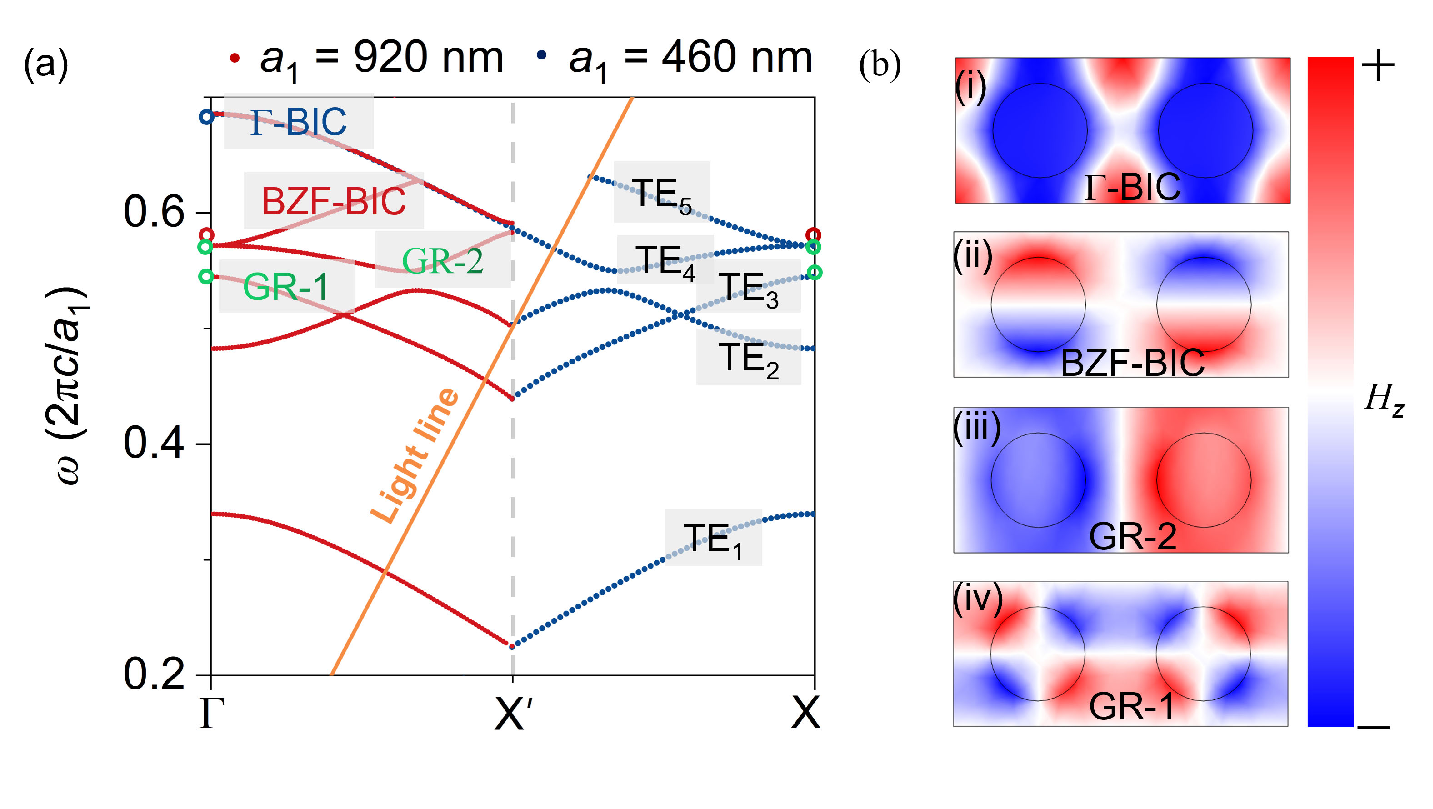}
\caption{\label{fig2} (a) Band diagrams of the transverse electric (TE) plane wave for both the unperturbed and gap-perturbed (asymmetry parameter $\alpha=0.025$) metasurfaces are illustrated using the blue and red dots, respectively. The GMs at the $X$ point of unperturbed metasurface are designated as TE$_m$ ($m=1$ to 5). When the GMs are folded into the $\Gamma$ point, the GRs, BZF-BIC, and $\Gamma$-BIC are indicated with green, red, and blue circles, respectively. (b) The magnetic field profiles $H_z$ of the above modes at $\Gamma$ point. The BICs and GRs exhibit even and odd features, respectively, under the C$_2$ operation around the $z$ axis.}
\end{figure}

 We employ an eigenfrequency solver to study the modal properties of the metasurface within the first Brillouin zone. Fig.~\ref{fig2} (a) presents the transverse electric (TE) band diagram along the $\Gamma-X$ direction for both the unperturbed and gap-perturbed metasurface (asymmetry parameter $\alpha=0.025$, the periodicity of the metasurface in the $x$ direction $a_{1}=920$ nm), plotted with blue and red solid dots, respectively. We observe five fundamental bands below the light line  (the origin solid line) are designated as TE$_{m}$ (where $m=1$ to 5), resulting in five modes at the $X$ point. Due to the total internal reflections, these modes are confined and localized in the transverse direction across the metasurface, termed the GMs. Upon introducing a gap perturbation ($\alpha=0.025$), these modes are folded into $\Gamma$ point, remaining above the light line. Notably, in addition to these five modes, there is another point at the $\Gamma$ point marked with a blue circle.  Considering only the realization of the third-order nonlinear optical response, we select four of these modes, indicated by different circles. Taking advantage of the modes’ symmetry corresponding to the $C_{2v}$ group, we can distinguish the GRs and BICs by calculating the out-of-plane components of their electromagnetic fields\cite{He2018, Overvig2020}. In Fig.~\ref{fig2}(b), we perform the magnetic field profiles $H_z$ of these modes at $\Gamma$ point. As shown, both the blue and red circles represent even modes under $C_2$ operation around the $z$ axis (see Figs.~\ref{fig2}(b-i) and ~\ref{fig2}(b-ii)), corresponding to the $\Gamma$-BIC and BZF-BIC respectively. The green circles exhibit opposite symmetries under $C_2$ operation (see Figs.~\ref{fig2}(b-iii) and~\ref{fig2}(b-iv), representing GRs.

\begin{figure*}[htbp]
\centering
\includegraphics
[scale=0.4]{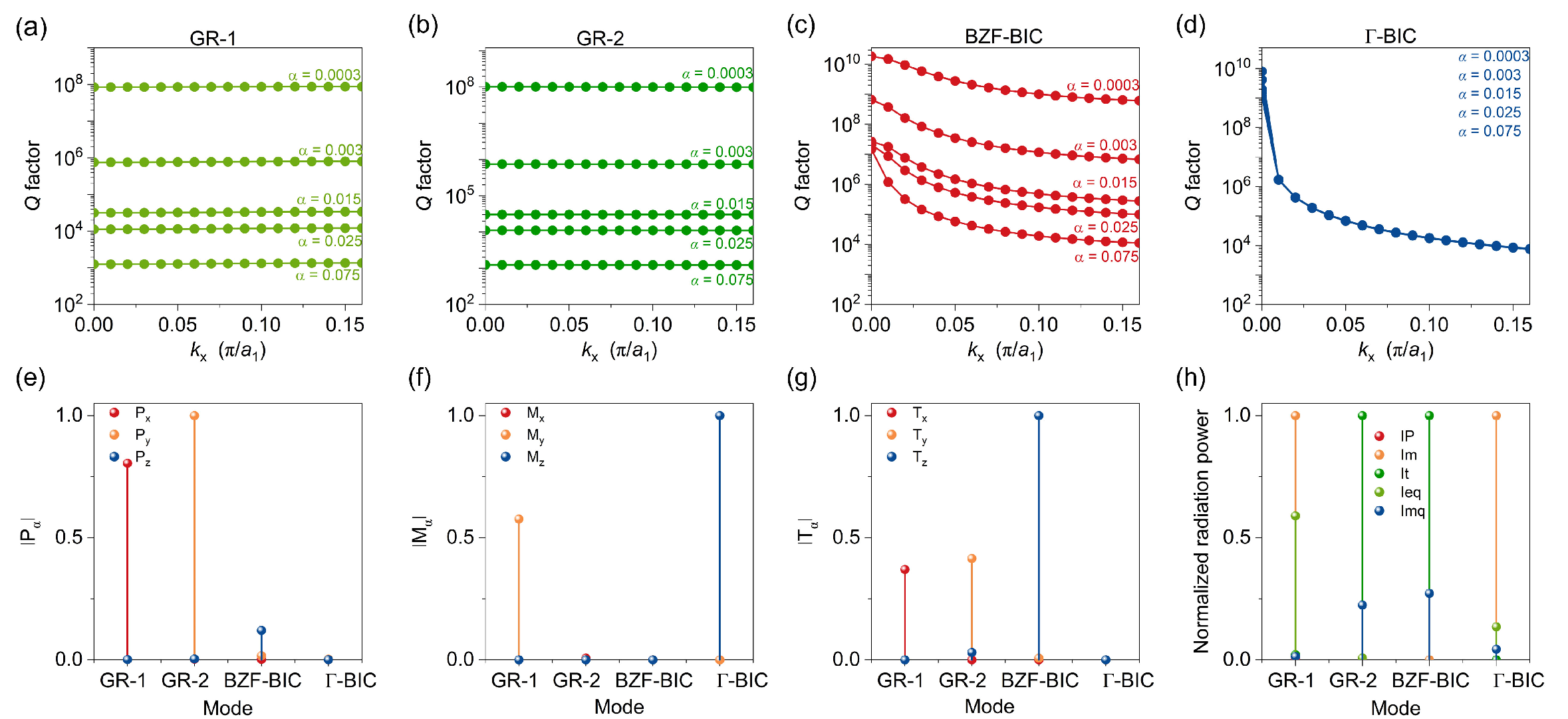}
\caption{\label{fig3} (a)-(d) Simulated $Q$ factors of GRs and BICs for different asymmetry parameters ($\alpha=0.0003$, 0.003, 0.015, 0.025, and 0.075). (e)-(g) Normalized multipole decomposition results of the eigenvalue for the GRs and BICs, with the components $|P_\alpha|$, $|M_\alpha|$, and $|T_\alpha|$ representing electric dipole (ED), magnetic dipole (MD), and toroidal dipole (TD), respectively. (h) Normalized radiation powers of the eigenvalue for the GRs and BICs.
}
\end{figure*}

To further analyze the excitation conditions of the GRs and BICs, we calculate the $Q$ factors and the induced Cartesian multipole decomposition as shown in Fig.~\ref{fig3}. From Figs.~\ref{fig3}(a)-\ref{fig3}(d), it is evident that the $Q$ factors of the two modes, GR-1 and GR-2, vary solely with the asymmetry parameter $\alpha$ and are independent of the wavevector $k_x$. In contrast, the $Q$ factor of the $\Gamma$-BIC exhibits the opposite trend, decreasing rapidly with increasing wavevector $k_x$ while remaining almost independent of the perturbation. The $Q$ factors of BZF-BIC, however, are influenced by both the asymmetry parameter $\alpha$ and the wavevector $k_x$ simultaneously. Therefore, we can draw the following conclusions in gap-perturbed metasurface: (i) although the GRs can theoretically be directly excited by a normally incident plane wave, it is necessary to increase the asymmetry parameter to obtain a clear visual representation in the transmission spectrum. (ii) the $\Gamma$-BIC is sufficient to simply increase the angle of the oblique incident wave to observe its transmission spectrum characteristics. (iii) the BZF-BIC requires simultaneous manipulation of both parameters to be effectively excited by the incident wave.

To more intuitively identify the excitation source of the eigenmode, we employ Cartesian multipole decomposition to quantitively identify the multipole character of GRs and BICs. In Fig.~\ref{fig3}(e), we observe that the GR-2 exhibits the largest $|P_y|$ component among all the modes, implying that it can be excited by an $E_y$-polarized normally incident plane wave. Conversely, the GR-1 mode shows a dominant $|P_x|$ component in Fig.~\ref{fig3}(e) and a prominent  $|M_y|$ component in Fig.~\ref{fig3}(f), suggesting that it can be excited by an  $E_x$-polarized normally incident plane wave. Notably, the BICs exhibit no multipole components that would contribute to far-field radiation along the $z$ direction, as shown in Figs.~\ref{fig3}(e)-~\ref{fig3}(f). This implies that these BICs can not be excited by a normally incident plane wave with any linear polarization, which aligns with the analysis of the magnetic field profiles and $Q$ factors presented above. At this point, it is important to determine the incident polarized for the excitation of these two BICs. 

It is known that whether an eigenmode can be excited depends on the symmetry matching condition between the eigenmode’ electric (magnetic) field profiles and the excitation source\cite{Lee2012, Wang2023}. By observing the magnetic field profiles of BICs in Fig.~\ref{fig2}(b), we find that the $\Gamma$-BIC exhibits an even feature, while the BZF-BIC exhibits odd feature under the $|\sigma_y|$ operation (change $y$ to $-y$). This behavior corresponds to the magnetic vector of $Ey$-polarized and $E_x$-polarized plane wave, respectively. Thus, we conclude that the $\Gamma$-BIC can be excited by an $E_y$-polarized plane wave, while the BZF-BIC mode can be excited by an $E_x$-polarized plane wave. In Fig.~\ref{fig3}(h), we present the normalized radiation powers of the eigenvalue for the GRs and BICs.  It can be seen that the GR-1 and $\Gamma$-BIC exhibit a toroidal dipole (TD) dominant, while the GR-2 and BZF-BIC exhibit a magnetic dipole (MD) dominant. Therefore, once the gap perturbation is induced, we can employ obliquely incident plane wave to excite these modes for different linear polarizations. 

\section{\label{sec:level3} Enhanced THG from nonlinear metasurfaces}

THG is defined as the transformation of an input wave at frequency $\omega$ into a new wave at frequency 3$\omega$. This nonlinear optical response can typically be analyzed using the frequency domain solver in COMSOL Multiphysics, and the process is divided into two main steps. First, the linear optical responses under $x/y$ -polarizations are computed as the fundamental frequency. Next, nonlinear polarization is introduced as a source for solving the wave equation at the harmonic frequency. For the realistic material of Si, the refractive index is derived from experimental data \cite{Palik2012}. The nonlinear polarization at 3$\omega$ can be expressed as\cite{Carletti2019, Boyd2020, Xiao2022}
\begin{equation}
	\textbf{\textit{P}}^{3\omega}=\varepsilon_{0}\chi^{(3)}(\textbf{\textit{E}}\cdot \textbf{\textit{E}})\textbf{\textit{E}}^\omega,
	\label{eq1}
\end{equation}
where $\varepsilon_0$ represents the vacuum permittivity, $\chi^{(3)}=2.45\times10^{-19}$ m$^2$/V$^2$ denotes the third-order nonlinear susceptibility of Si in the near-infrared region, $\textbf{\textit{E}}^\omega$ is the electric field of the fundamental frequency $\omega$, and the input optical intensity of the fundamental frequency is fixed at 1 MW/cm$^2$. Based on the excitation conditions previously obtained, we can explore the THG response under $x$ and $y$-polarization.

\begin{figure}
[htbp]
\centering
\includegraphics
[scale=0.6]{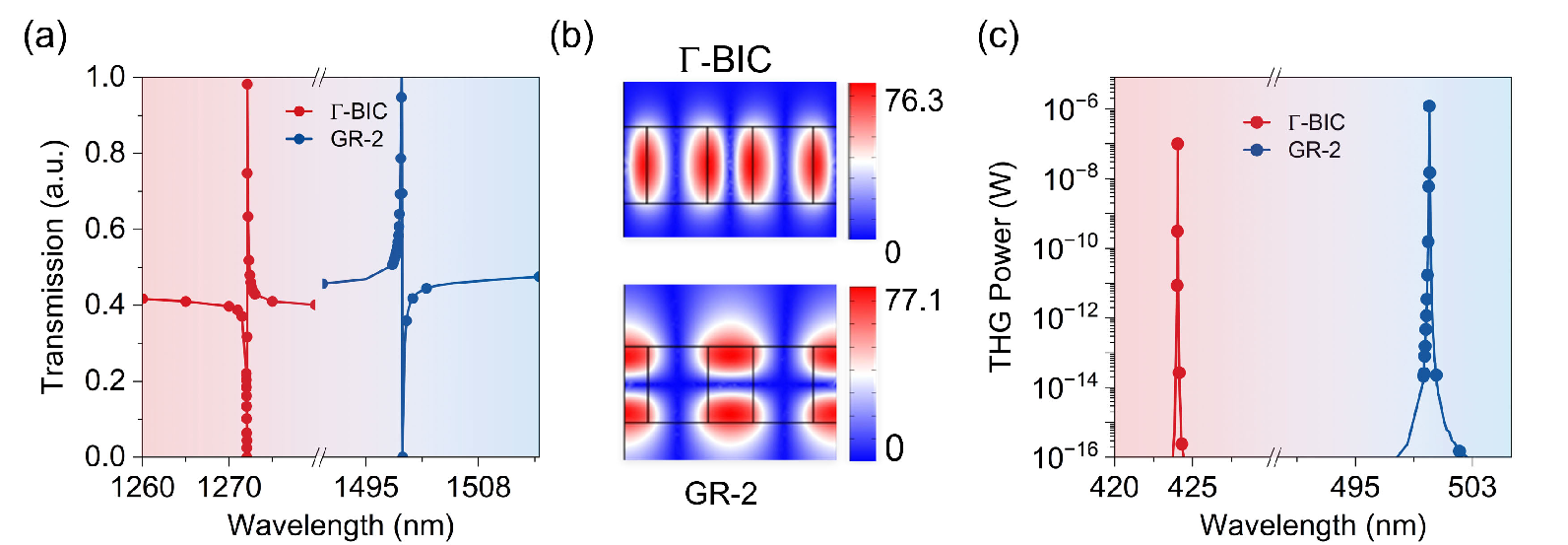}
\caption{\label{fig4}Simulated the optical response of the $y$-polarized plane wave under an oblique incident angle of $1^\circ$ for $\Gamma$-BIC (red) and GR-2  (blue), with an asymmetry parameter $\alpha=0.025$. (a) The linear transmission spectra of the two resonance modes. (b) The electric field enhancement in the ($x,z$) plane at incident wavelength of 1272.08 nm ($\Gamma$-BIC) and 1499.24 nm (GR-2), respectively. (c) The output power of the THG for the $\Gamma$-BIC (red) and  GR-2 (blue).}
\end{figure}

As discussed previously, the introduction of gap perturbations allows for the simultaneous exciton of the GR-2 and the $\Gamma$-BIC by obliquely incident $y$ -polarized waves. Therefore, we will next examine the enhancement of THG associated with these two modes under $y$-polarized light. With the asymmetry parameter $\alpha=0.025$ and the oblique incident angle at $1^\circ$, we can clearly observe two resonance modes with the locations are 1272.08 nm and 1499.24 nm appeared in Fig.~\ref{fig4}(a), they are respectively  $\Gamma$-BIC (red) and GR-2 (blue). A comparison of the electric field enhancement in the $(x,z)$ plane for both modes, as shown in Fig.~\ref{fig4}(b), reveals that the electric field intensity of the GR-2 is slightly higher than that of the $\Gamma$-BIC. Consequently, as illustrated in Fig.~\ref{fig4}(c), the output power of the GR-2 is also marginally higher than that of the $\Gamma$-BIC. This observation indicates that the enhancement of the THG is closely related to the electric field enhancement associated with the resonant modes.

\begin{figure}[htbp]
\centering
\includegraphics
[scale=0.6]{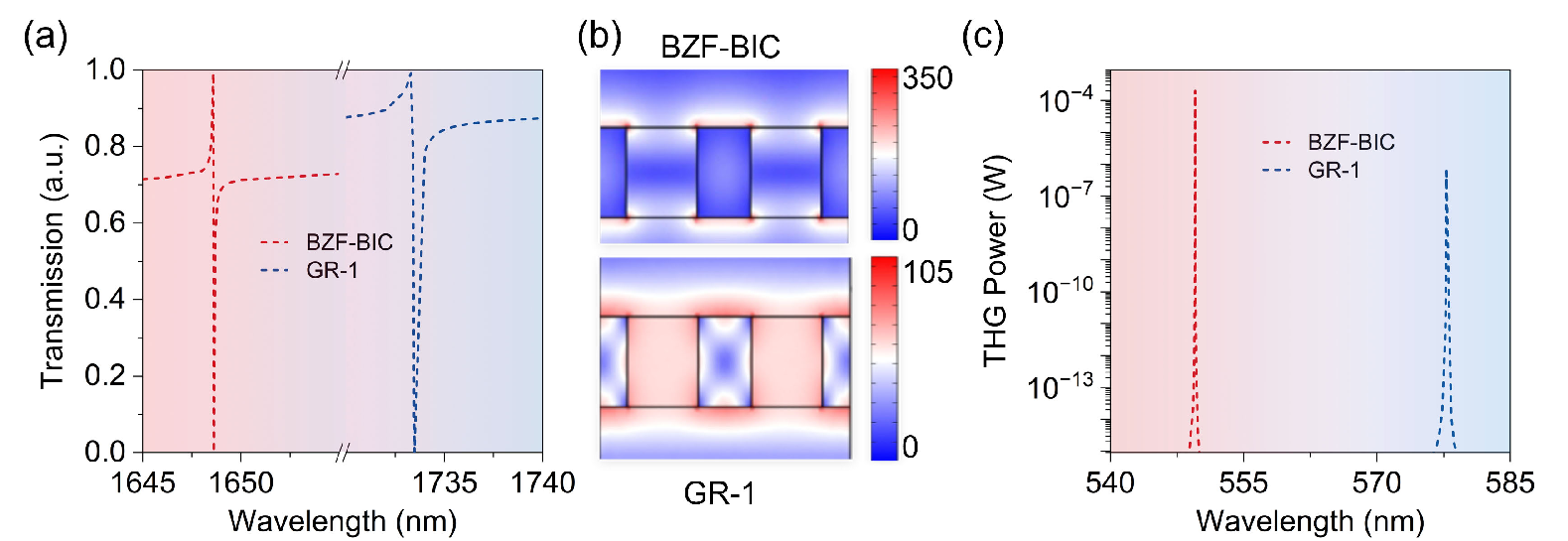}
\caption{\label{fig5}Simulated the optical response of the $x$-polarized plane wave under an oblique incident angle of $5^\circ$ for the BZF-BIC (red) and GR-1 (blue), with an asymmetry parameter $\alpha=0.025$. The linear transmission spectra of the two resonance modes. (b) The electric field enhancement in the $(x,y)$ plane at incident wavelength of 1648.614 nm (BZF-BIC) and 1733.49 nm (GR-1), respectively.
(c) The output power of the THG for the BZF-BIC (red) and GR-1 (blue).}
\end{figure}

To simultaneously excite the GR-1 and the BZF-BIC, we set the asymmetry parameter $\alpha=0.025$ and the incident angle of the $x$-polarized light to $5^\circ$. In Fig.~\ref{fig5}(a), two resonance modes can be identified at wavelengths of 1648.614 nm and 1733.49 nm. The red dashed line, representing the BZF-BIC, exhibits a narrower linewidth, while the blue dashed line corresponds to the GR-1. This can be further elucidated by Fig.~\ref{fig5}(b) which shows that the local electric field enhancement for the BZF-BIC reached a value of 350, significantly surpassing the enhancements observed in the other three modes. This substantial enhancement results in a THG output power exceeding $10^{-4}$ W, which further indicate that the BZF-BIC mode represents an ideal pathway for enhancing THG.

\section{\label{sec:level4}Enhanced dFWM from nonlinear metasurfaces}
\begin{figure}[htbp]
\centering
\includegraphics
[scale=0.7]{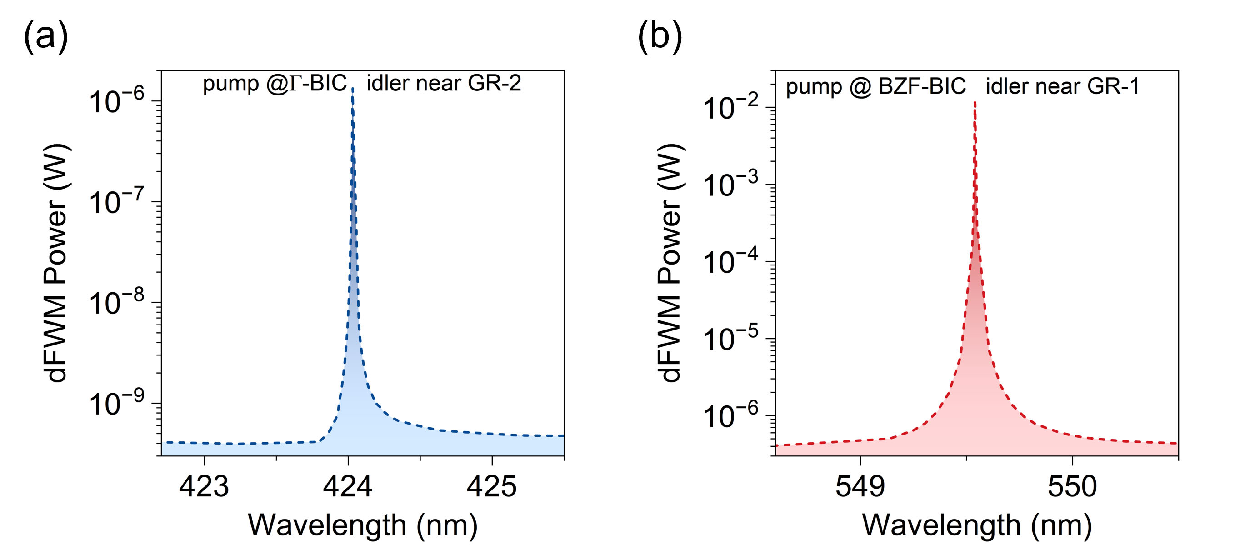}
\caption{\label{fig6} Enhanced third-order optical nonlinearities through dFWM. (a) Output power of dWFM as a function of the idler wavelength of the GR-2, with the pump wavelength fixed at the $\Gamma$-BIC. (b) Output power of dWFM as a function of the idler wavelength of the GR-1, with the pump wavelength is fixed at the BZF-BIC.}
\end{figure}
Four-wave mixing (FWM) is defined as a process in which three photons of different frequencies are mixed to generate a new frequency. When the two input pump photons have equal energy, FWM  transforms into dFWM. In this case, two photons with the same frequency $\omega_1$ serve as the pump, while another photon with a different frequency $\omega_2$ acts as the idler. The generated new frequency $\omega_3$ can be expressed as: $\omega_3=2\omega_1+\omega_2$, with the polarization terms satisfying the relation\cite{Boyd2020,Carletti2019,Xiao2022}

\begin{equation}
   \textbf{\textit{P}}^{3\omega}= 3{\varepsilon _0}{\chi ^{(3)}}\left[ {2({\textbf{\textit{E}}^{{\omega _1}}} \cdot {\textbf{\textit{E}}^{{\omega _2}}}){\textbf{\textit{E}}^{{\omega _1}}} + ({\textbf{\textit{E}}^{{\omega _1}}} \cdot {\textbf{\textit{E}}^{{\omega _1}}}){\textbf{\textit{E}}^{{\omega _2}}}} \right],
	\label{eq2}  
\end{equation}

where $\textbf{\textit{E}}^{{\omega _1}}$ and $\textbf{\textit{E}}^{{\omega _2}}$ represent the electric fields at frequency $\omega_1$ and $\omega_2$, respectively. For easy comparisons, the input pumps are also fixed at 1 MW/cm$^2$ in the subsequent simulations. In Fig.~\ref{fig6}(a), the $\Gamma$-BIC is designated as the pump while the GR-2 mode serves as the idler. It is readily apparent that the output power of the dFWM near 424 nm reaches 10$^{-6}$ W, which is clearly one order of magnitude higher than the THG solely enhanced by the $\Gamma$-BIC. This enhancement can be attributed to the overlap of the two modes, resulting in a $Q$ factor that is proportional to the product of the $Q$ factors of the original two modes, thereby significantly enhancing the electric local field. A similar phenomenon is observed in Fig.~\ref{fig6}(b), where the BZF-BIC mode is designated as the pump and the GR-1 mode as the idler. In this case, the output power of dFWM near 549.5 nm exceeds 10$^{-2}$ W, which is two orders of magnitude higher than the THG enhancement contributed by the BZF-BIC alone.

\section{\label{sec:level3} CONCLUSIONS}

In conclusion, we achieve effective enhancement of THG and dFWM in an air-hole-type nonlinear metasurfaces under dual polarization. By analyzing the $Q$ factors and the electromagnetic radiation characteristics of the eigenmodes, we establish the excitation conditions for the GRs, $\Gamma$-BIC, and BZF-BIC, thereby laying a theoretical foundation for the subsequent enhancement of third-order nonlinearity. Comparing the performance of these modes in enhancing THG and dFWM, we found that the BZF-BIC mode demonstrates a significant improvement in third-order nonlinearity. Under an input power density of 1  MW/cm$^2$, the output power of THG reaches 10$^{-4}$ W, and the outpower of dFWM can exceed 10$^{-2}$ W. Our work offers a simple but effective approach for enhancing the nonlinear optical response, significantly broadening the development and application of novel functional optoelectronic devices.

\begin{acknowledgments}
This work is supported by the National Natural Science Foundation of China (Grants No. 12364049, No. 12304420, No. 12264028, No. 12364045,  and No. 12104105),  the Natural Science Foundation of Jiangxi Province (Grants No. 20232BAB211025, No. 20232BAB201040, and No. 20242BAB25041), the Guangdong Basic and Applied Basic Research Foundation (Grant No. 2023A1515011024), the Start-up Funding of Nanchang Institute of Science and Technology (Grant No. NGRCZX-23–01), and the Scientific research project of  Nanchang  Institute of Science and Technology (Grant No. NGKJ-23-02), the Young Elite Scientists Sponsorship Program by JXAST (Grants No. 2023QT11 and 2025QT04), and the Innovation Fund for Graduate Students of Jiangxi Province (Grant No. YC2022-B019).
\end{acknowledgments}

\nocite{*}


\providecommand{\noopsort}[1]{}\providecommand{\singleletter}[1]{#1}%
\begin{thebibliography}{58}%
\makeatletter
\providecommand \@ifxundefined [1]{%
 \@ifx{#1\undefined}
}%
\providecommand \@ifnum [1]{%
 \ifnum #1\expandafter \@firstoftwo
 \else \expandafter \@secondoftwo
 \fi
}%
\providecommand \@ifx [1]{%
 \ifx #1\expandafter \@firstoftwo
 \else \expandafter \@secondoftwo
 \fi
}%
\providecommand \natexlab [1]{#1}%
\providecommand \enquote  [1]{``#1''}%
\providecommand \bibnamefont  [1]{#1}%
\providecommand \bibfnamefont [1]{#1}%
\providecommand \citenamefont [1]{#1}%
\providecommand \href@noop [0]{\@secondoftwo}%
\providecommand \href [0]{\begingroup \@sanitize@url \@href}%
\providecommand \@href[1]{\@@startlink{#1}\@@href}%
\providecommand \@@href[1]{\endgroup#1\@@endlink}%
\providecommand \@sanitize@url [0]{\catcode `\\12\catcode `\$12\catcode
  `\&12\catcode `\#12\catcode `\^12\catcode `\_12\catcode `\%12\relax}%
\providecommand \@@startlink[1]{}%
\providecommand \@@endlink[0]{}%
\providecommand \url  [0]{\begingroup\@sanitize@url \@url }%
\providecommand \@url [1]{\endgroup\@href {#1}{\urlprefix }}%
\providecommand \urlprefix  [0]{URL }%
\providecommand \Eprint [0]{\href }%
\providecommand \doibase [0]{https://doi.org/}%
\providecommand \selectlanguage [0]{\@gobble}%
\providecommand \bibinfo  [0]{\@secondoftwo}%
\providecommand \bibfield  [0]{\@secondoftwo}%
\providecommand \translation [1]{[#1]}%
\providecommand \BibitemOpen [0]{}%
\providecommand \bibitemStop [0]{}%
\providecommand \bibitemNoStop [0]{.\EOS\space}%
\providecommand \EOS [0]{\spacefactor3000\relax}%
\providecommand \BibitemShut  [1]{\csname bibitem#1\endcsname}%
\let\auto@bib@innerbib\@empty
\bibitem [{\citenamefont {R.W.Boyd}(2020)}]{Boyd2020}%
  \BibitemOpen
  \bibfield  {author} {\bibinfo {author} {\bibnamefont {R.W.Boyd}},\ }\href
  {https://books.google.com/books?id=54vZDwAAQBAJ} {\emph {\bibinfo {title}
  {Nonlinear Optics}}}\ (\bibinfo  {publisher} {Academic Press},\ \bibinfo
  {year} {2020})\BibitemShut {NoStop}%
\bibitem [{\citenamefont {Sheik-Bahae}\ and\ \citenamefont
  {Hasselbeck}(2000)}]{sheik2000third}%
  \BibitemOpen
  \bibfield  {author} {\bibinfo {author} {\bibfnamefont {M.}~\bibnamefont
  {Sheik-Bahae}}\ and\ \bibinfo {author} {\bibfnamefont {M.~P.}\ \bibnamefont
  {Hasselbeck}},\ }\bibfield  {title} {\bibinfo {title} {Third-order optical
  nonlinearities},\ }\href@noop {} {\bibfield  {journal} {\bibinfo  {journal}
  {Handbook of Optics}\ }\textbf {\bibinfo {volume} {4}},\ \bibinfo {pages}
  {16} (\bibinfo {year} {2000})}\BibitemShut {NoStop}%
\bibitem [{\citenamefont {Dinu}\ \emph {et~al.}(2003)\citenamefont {Dinu},
  \citenamefont {Quochi},\ and\ \citenamefont {Garcia}}]{dinu2003third}%
  \BibitemOpen
  \bibfield  {author} {\bibinfo {author} {\bibfnamefont {M.}~\bibnamefont
  {Dinu}}, \bibinfo {author} {\bibfnamefont {F.}~\bibnamefont {Quochi}},\ and\
  \bibinfo {author} {\bibfnamefont {H.}~\bibnamefont {Garcia}},\ }\bibfield
  {title} {\bibinfo {title} {Third-order nonlinearities in silicon at telecom
  wavelengths},\ }\href@noop {} {\bibfield  {journal} {\bibinfo  {journal}
  {Applied Physics Letters}\ }\textbf {\bibinfo {volume} {82}},\ \bibinfo
  {pages} {2954} (\bibinfo {year} {2003})}\BibitemShut {NoStop}%
\bibitem [{\citenamefont {Hutchinson}\ and\ \citenamefont
  {Milburn}(2004)}]{Hutchinson2004}%
  \BibitemOpen
  \bibfield  {author} {\bibinfo {author} {\bibfnamefont {G.~D.}\ \bibnamefont
  {Hutchinson}}\ and\ \bibinfo {author} {\bibfnamefont {G.~J.}\ \bibnamefont
  {Milburn}},\ }\bibfield  {title} {\bibinfo {title} {Nonlinear quantum optical
  computing via measurement},\ }\href
  {https://doi.org/10.1080/09500340408230417} {\bibfield  {journal} {\bibinfo
  {journal} {Journal of Modern Optics}\ }\textbf {\bibinfo {volume} {51}},\
  \bibinfo {pages} {1211} (\bibinfo {year} {2004})}\BibitemShut {NoStop}%
\bibitem [{\citenamefont {Xu}\ \emph {et~al.}(2019)\citenamefont {Xu},
  \citenamefont {Zangeneh~Kamali}, \citenamefont {Huang}, \citenamefont
  {Rahmani}, \citenamefont {Smirnov}, \citenamefont {Camacho-Morales},
  \citenamefont {Ma}, \citenamefont {Zhang}, \citenamefont {Woolley},
  \citenamefont {Neshev} \emph {et~al.}}]{Xu2019}%
  \BibitemOpen
  \bibfield  {author} {\bibinfo {author} {\bibfnamefont {L.}~\bibnamefont
  {Xu}}, \bibinfo {author} {\bibfnamefont {K.}~\bibnamefont {Zangeneh~Kamali}},
  \bibinfo {author} {\bibfnamefont {L.}~\bibnamefont {Huang}}, \bibinfo
  {author} {\bibfnamefont {M.}~\bibnamefont {Rahmani}}, \bibinfo {author}
  {\bibfnamefont {A.}~\bibnamefont {Smirnov}}, \bibinfo {author} {\bibfnamefont
  {R.}~\bibnamefont {Camacho-Morales}}, \bibinfo {author} {\bibfnamefont
  {Y.}~\bibnamefont {Ma}}, \bibinfo {author} {\bibfnamefont {G.}~\bibnamefont
  {Zhang}}, \bibinfo {author} {\bibfnamefont {M.}~\bibnamefont {Woolley}},
  \bibinfo {author} {\bibfnamefont {D.}~\bibnamefont {Neshev}}, \emph
  {et~al.},\ }\bibfield  {title} {\bibinfo {title} {Dynamic nonlinear image
  tuning through magnetic dipole quasi-bic ultrathin resonators},\ }\href@noop
  {} {\bibfield  {journal} {\bibinfo  {journal} {Advanced science}\ }\textbf
  {\bibinfo {volume} {6}},\ \bibinfo {pages} {1802119} (\bibinfo {year}
  {2019})}\BibitemShut {NoStop}%
\bibitem [{\citenamefont {Gao}\ \emph {et~al.}(2018)\citenamefont {Gao},
  \citenamefont {Fan}, \citenamefont {Wang}, \citenamefont {Yang},
  \citenamefont {Song},\ and\ \citenamefont {Xiao}}]{gao2018nonlinear}%
  \BibitemOpen
  \bibfield  {author} {\bibinfo {author} {\bibfnamefont {Y.}~\bibnamefont
  {Gao}}, \bibinfo {author} {\bibfnamefont {Y.}~\bibnamefont {Fan}}, \bibinfo
  {author} {\bibfnamefont {Y.}~\bibnamefont {Wang}}, \bibinfo {author}
  {\bibfnamefont {W.}~\bibnamefont {Yang}}, \bibinfo {author} {\bibfnamefont
  {Q.}~\bibnamefont {Song}},\ and\ \bibinfo {author} {\bibfnamefont
  {S.}~\bibnamefont {Xiao}},\ }\bibfield  {title} {\bibinfo {title} {Nonlinear
  holographic all-dielectric metasurfaces},\ }\href@noop {} {\bibfield
  {journal} {\bibinfo  {journal} {Nano letters}\ }\textbf {\bibinfo {volume}
  {18}},\ \bibinfo {pages} {8054} (\bibinfo {year} {2018})}\BibitemShut
  {NoStop}%
\bibitem [{\citenamefont {Maker}\ \emph {et~al.}(1962)\citenamefont {Maker},
  \citenamefont {Terhune}, \citenamefont {Nisenoff},\ and\ \citenamefont
  {Savage}}]{Maker1962}%
  \BibitemOpen
  \bibfield  {author} {\bibinfo {author} {\bibfnamefont {P.~D.}\ \bibnamefont
  {Maker}}, \bibinfo {author} {\bibfnamefont {R.~W.}\ \bibnamefont {Terhune}},
  \bibinfo {author} {\bibfnamefont {M.}~\bibnamefont {Nisenoff}},\ and\
  \bibinfo {author} {\bibfnamefont {C.~M.}\ \bibnamefont {Savage}},\ }\bibfield
   {title} {\bibinfo {title} {Effects of dispersion and focusing on the
  production of optical harmonics},\ }\href
  {https://doi.org/10.1103/physrevlett.8.21} {\bibfield  {journal} {\bibinfo
  {journal} {Physical Review Letters}\ }\textbf {\bibinfo {volume} {8}},\
  \bibinfo {pages} {21} (\bibinfo {year} {1962})}\BibitemShut {NoStop}%
\bibitem [{\citenamefont {Lifshitz}\ \emph {et~al.}(2005)\citenamefont
  {Lifshitz}, \citenamefont {Arie},\ and\ \citenamefont
  {Bahabad}}]{Lifshitz2005}%
  \BibitemOpen
  \bibfield  {author} {\bibinfo {author} {\bibfnamefont {R.}~\bibnamefont
  {Lifshitz}}, \bibinfo {author} {\bibfnamefont {A.}~\bibnamefont {Arie}},\
  and\ \bibinfo {author} {\bibfnamefont {A.}~\bibnamefont {Bahabad}},\
  }\bibfield  {title} {\bibinfo {title} {Photonic quasicrystals for nonlinear
  optical frequency conversion},\ }\href
  {https://doi.org/10.1103/physrevlett.95.133901} {\bibfield  {journal}
  {\bibinfo  {journal} {Physical Review Letters}\ }\textbf {\bibinfo {volume}
  {95}},\ \bibinfo {pages} {133901} (\bibinfo {year} {2005})}\BibitemShut
  {NoStop}%
\bibitem [{\citenamefont {Paul}\ \emph {et~al.}(2003)\citenamefont {Paul},
  \citenamefont {Bartels}, \citenamefont {Tobey}, \citenamefont {Green},
  \citenamefont {Weiman}, \citenamefont {Christov}, \citenamefont {Murnane},
  \citenamefont {Kapteyn},\ and\ \citenamefont {Backus}}]{Paul2003}%
  \BibitemOpen
  \bibfield  {author} {\bibinfo {author} {\bibfnamefont {A.}~\bibnamefont
  {Paul}}, \bibinfo {author} {\bibfnamefont {R.~A.}\ \bibnamefont {Bartels}},
  \bibinfo {author} {\bibfnamefont {R.}~\bibnamefont {Tobey}}, \bibinfo
  {author} {\bibfnamefont {H.}~\bibnamefont {Green}}, \bibinfo {author}
  {\bibfnamefont {S.}~\bibnamefont {Weiman}}, \bibinfo {author} {\bibfnamefont
  {I.~P.}\ \bibnamefont {Christov}}, \bibinfo {author} {\bibfnamefont {M.~M.}\
  \bibnamefont {Murnane}}, \bibinfo {author} {\bibfnamefont {H.~C.}\
  \bibnamefont {Kapteyn}},\ and\ \bibinfo {author} {\bibfnamefont
  {S.}~\bibnamefont {Backus}},\ }\bibfield  {title} {\bibinfo {title}
  {Quasi-phase-matched generation of coherent extreme-ultraviolet light},\
  }\href {https://doi.org/10.1038/nature01222} {\bibfield  {journal} {\bibinfo
  {journal} {Nature}\ }\textbf {\bibinfo {volume} {421}},\ \bibinfo {pages}
  {51} (\bibinfo {year} {2003})}\BibitemShut {NoStop}%
\bibitem [{\citenamefont {Krasnok}\ \emph {et~al.}(2018)\citenamefont
  {Krasnok}, \citenamefont {Tymchenko},\ and\ \citenamefont
  {Alù}}]{Krasnok2018}%
  \BibitemOpen
  \bibfield  {author} {\bibinfo {author} {\bibfnamefont {A.}~\bibnamefont
  {Krasnok}}, \bibinfo {author} {\bibfnamefont {M.}~\bibnamefont {Tymchenko}},\
  and\ \bibinfo {author} {\bibfnamefont {A.}~\bibnamefont {Alù}},\ }\bibfield
  {title} {\bibinfo {title} {Nonlinear metasurfaces: a paradigm shift in
  nonlinear optics},\ }\href {https://doi.org/10.1016/j.mattod.2017.06.007}
  {\bibfield  {journal} {\bibinfo  {journal} {Materials Today}\ }\textbf
  {\bibinfo {volume} {21}},\ \bibinfo {pages} {8} (\bibinfo {year}
  {2018})}\BibitemShut {NoStop}%
\bibitem [{\citenamefont {Huang}\ \emph {et~al.}(2020)\citenamefont {Huang},
  \citenamefont {Zhao}, \citenamefont {Zeng}, \citenamefont {Crunteanu},
  \citenamefont {Shum},\ and\ \citenamefont {Yu}}]{Huang2020}%
  \BibitemOpen
  \bibfield  {author} {\bibinfo {author} {\bibfnamefont {T.}~\bibnamefont
  {Huang}}, \bibinfo {author} {\bibfnamefont {X.}~\bibnamefont {Zhao}},
  \bibinfo {author} {\bibfnamefont {S.}~\bibnamefont {Zeng}}, \bibinfo {author}
  {\bibfnamefont {A.}~\bibnamefont {Crunteanu}}, \bibinfo {author}
  {\bibfnamefont {P.~P.}\ \bibnamefont {Shum}},\ and\ \bibinfo {author}
  {\bibfnamefont {N.}~\bibnamefont {Yu}},\ }\bibfield  {title} {\bibinfo
  {title} {Planar nonlinear metasurface optics and their applications},\ }\href
  {https://doi.org/10.1088/1361-6633/abb56e} {\bibfield  {journal} {\bibinfo
  {journal} {Reports on Progress in Physics}\ }\textbf {\bibinfo {volume}
  {83}},\ \bibinfo {pages} {126101} (\bibinfo {year} {2020})}\BibitemShut
  {NoStop}%
\bibitem [{\citenamefont {Kivshar}(2018)}]{kivshar2018all}%
  \BibitemOpen
  \bibfield  {author} {\bibinfo {author} {\bibfnamefont {Y.}~\bibnamefont
  {Kivshar}},\ }\bibfield  {title} {\bibinfo {title} {All-dielectric
  meta-optics and non-linear nanophotonics},\ }\href@noop {} {\bibfield
  {journal} {\bibinfo  {journal} {National Science Review}\ }\textbf {\bibinfo
  {volume} {5}},\ \bibinfo {pages} {144} (\bibinfo {year} {2018})}\BibitemShut
  {NoStop}%
\bibitem [{\citenamefont {Liu}\ \emph {et~al.}(2022)\citenamefont {Liu},
  \citenamefont {Xiao}, \citenamefont {Li}, \citenamefont {Gu}, \citenamefont
  {Luan},\ and\ \citenamefont {Fang}}]{Liu2022}%
  \BibitemOpen
  \bibfield  {author} {\bibinfo {author} {\bibfnamefont {T.}~\bibnamefont
  {Liu}}, \bibinfo {author} {\bibfnamefont {S.}~\bibnamefont {Xiao}}, \bibinfo
  {author} {\bibfnamefont {B.}~\bibnamefont {Li}}, \bibinfo {author}
  {\bibfnamefont {M.}~\bibnamefont {Gu}}, \bibinfo {author} {\bibfnamefont
  {H.}~\bibnamefont {Luan}},\ and\ \bibinfo {author} {\bibfnamefont
  {X.}~\bibnamefont {Fang}},\ }\bibfield  {title} {\bibinfo {title} {Third-and
  second-harmonic generation in all-dielectric nanostructures: a mini review},\
  }\href@noop {} {\bibfield  {journal} {\bibinfo  {journal} {Frontiers in
  Nanotechnology}\ }\textbf {\bibinfo {volume} {4}},\ \bibinfo {pages} {891892}
  (\bibinfo {year} {2022})}\BibitemShut {NoStop}%
\bibitem [{\citenamefont {Vabishchevich}\ and\ \citenamefont
  {Kivshar}(2023)}]{Vabishchevich2023}%
  \BibitemOpen
  \bibfield  {author} {\bibinfo {author} {\bibfnamefont {P.}~\bibnamefont
  {Vabishchevich}}\ and\ \bibinfo {author} {\bibfnamefont {Y.}~\bibnamefont
  {Kivshar}},\ }\bibfield  {title} {\bibinfo {title} {Nonlinear photonics with
  metasurfaces},\ }\href {https://doi.org/10.1364/prj.474387} {\bibfield
  {journal} {\bibinfo  {journal} {Photonics Research}\ }\textbf {\bibinfo
  {volume} {11}},\ \bibinfo {pages} {B50} (\bibinfo {year} {2023})}\BibitemShut
  {NoStop}%
\bibitem [{\citenamefont {Xu}\ \emph {et~al.}(2024)\citenamefont {Xu},
  \citenamefont {Ma}, \citenamefont {Zhu}, \citenamefont {Dong}, \citenamefont
  {Jiang}, \citenamefont {Gao}, \citenamefont {Qin}, \citenamefont {Liu},
  \citenamefont {Wu},\ and\ \citenamefont {Li}}]{Xu2024}%
  \BibitemOpen
  \bibfield  {author} {\bibinfo {author} {\bibfnamefont {H.}~\bibnamefont
  {Xu}}, \bibinfo {author} {\bibfnamefont {R.}~\bibnamefont {Ma}}, \bibinfo
  {author} {\bibfnamefont {Y.}~\bibnamefont {Zhu}}, \bibinfo {author}
  {\bibfnamefont {A.}~\bibnamefont {Dong}}, \bibinfo {author} {\bibfnamefont
  {H.}~\bibnamefont {Jiang}}, \bibinfo {author} {\bibfnamefont
  {W.}~\bibnamefont {Gao}}, \bibinfo {author} {\bibfnamefont {M.}~\bibnamefont
  {Qin}}, \bibinfo {author} {\bibfnamefont {J.}~\bibnamefont {Liu}}, \bibinfo
  {author} {\bibfnamefont {F.}~\bibnamefont {Wu}},\ and\ \bibinfo {author}
  {\bibfnamefont {H.}~\bibnamefont {Li}},\ }\bibfield  {title} {\bibinfo
  {title} {Chirality-controlled second-order nonlinear frequency conversion in
  lithium niobate film metasurfaces},\ }\href
  {https://doi.org/10.1364/ol.538625} {\bibfield  {journal} {\bibinfo
  {journal} {Optics Letters}\ }\textbf {\bibinfo {volume} {49}},\ \bibinfo
  {pages} {6405} (\bibinfo {year} {2024})}\BibitemShut {NoStop}%
\bibitem [{\citenamefont {Wang}\ \emph {et~al.}(2024)\citenamefont {Wang},
  \citenamefont {Tonkaev}, \citenamefont {Koshelev}, \citenamefont {Lai},
  \citenamefont {Kruk}, \citenamefont {Song}, \citenamefont {Kivshar},\ and\
  \citenamefont {Panoiu}}]{Wang2024}%
  \BibitemOpen
  \bibfield  {author} {\bibinfo {author} {\bibfnamefont {J.~T.}\ \bibnamefont
  {Wang}}, \bibinfo {author} {\bibfnamefont {P.}~\bibnamefont {Tonkaev}},
  \bibinfo {author} {\bibfnamefont {K.}~\bibnamefont {Koshelev}}, \bibinfo
  {author} {\bibfnamefont {F.}~\bibnamefont {Lai}}, \bibinfo {author}
  {\bibfnamefont {S.}~\bibnamefont {Kruk}}, \bibinfo {author} {\bibfnamefont
  {Q.}~\bibnamefont {Song}}, \bibinfo {author} {\bibfnamefont {Y.}~\bibnamefont
  {Kivshar}},\ and\ \bibinfo {author} {\bibfnamefont {N.~C.}\ \bibnamefont
  {Panoiu}},\ }\bibfield  {title} {\bibinfo {title} {Resonantly enhanced
  second-and third-harmonic generation in dielectric nonlinear metasurfaces},\
  }\href {https://doi.org/10.29026/oea.2024.230186} {\bibfield  {journal}
  {\bibinfo  {journal} {Opto-Electronic Advances}\ }\textbf {\bibinfo {volume}
  {7}},\ \bibinfo {pages} {230186} (\bibinfo {year} {2024})}\BibitemShut
  {NoStop}%
\bibitem [{\citenamefont {Panoiu}\ \emph {et~al.}(2018)\citenamefont {Panoiu},
  \citenamefont {Sha}, \citenamefont {Lei},\ and\ \citenamefont
  {Li}}]{Panoiu2018}%
  \BibitemOpen
  \bibfield  {author} {\bibinfo {author} {\bibfnamefont {N.~C.}\ \bibnamefont
  {Panoiu}}, \bibinfo {author} {\bibfnamefont {W.~E.~I.}\ \bibnamefont {Sha}},
  \bibinfo {author} {\bibfnamefont {D.~Y.}\ \bibnamefont {Lei}},\ and\ \bibinfo
  {author} {\bibfnamefont {G.-C.}\ \bibnamefont {Li}},\ }\bibfield  {title}
  {\bibinfo {title} {Nonlinear optics in plasmonic nanostructures},\ }\href
  {https://doi.org/10.1088/2040-8986/aac8ed} {\bibfield  {journal} {\bibinfo
  {journal} {Journal of Optics}\ }\textbf {\bibinfo {volume} {20}},\ \bibinfo
  {pages} {083001} (\bibinfo {year} {2018})}\BibitemShut {NoStop}%
\bibitem [{\citenamefont {Bin-Alam}\ \emph {et~al.}(2021)\citenamefont
  {Bin-Alam}, \citenamefont {Reshef}, \citenamefont {Mamchur}, \citenamefont
  {Alam}, \citenamefont {Carlow}, \citenamefont {Upham}, \citenamefont
  {Sullivan}, \citenamefont {Ménard}, \citenamefont {Huttunen}, \citenamefont
  {Boyd},\ and\ \citenamefont {Dolgaleva}}]{BinAlam2021}%
  \BibitemOpen
  \bibfield  {author} {\bibinfo {author} {\bibfnamefont {M.~S.}\ \bibnamefont
  {Bin-Alam}}, \bibinfo {author} {\bibfnamefont {O.}~\bibnamefont {Reshef}},
  \bibinfo {author} {\bibfnamefont {Y.}~\bibnamefont {Mamchur}}, \bibinfo
  {author} {\bibfnamefont {M.~Z.}\ \bibnamefont {Alam}}, \bibinfo {author}
  {\bibfnamefont {G.}~\bibnamefont {Carlow}}, \bibinfo {author} {\bibfnamefont
  {J.}~\bibnamefont {Upham}}, \bibinfo {author} {\bibfnamefont {B.~T.}\
  \bibnamefont {Sullivan}}, \bibinfo {author} {\bibfnamefont {J.-M.}\
  \bibnamefont {Ménard}}, \bibinfo {author} {\bibfnamefont {M.~J.}\
  \bibnamefont {Huttunen}}, \bibinfo {author} {\bibfnamefont {R.~W.}\
  \bibnamefont {Boyd}},\ and\ \bibinfo {author} {\bibfnamefont
  {K.}~\bibnamefont {Dolgaleva}},\ }\bibfield  {title} {\bibinfo {title}
  {Ultra-high-q resonances in plasmonic metasurfaces},\ }\href
  {https://doi.org/10.1038/s41467-021-21196-2} {\bibfield  {journal} {\bibinfo
  {journal} {Nature Communications}\ }\textbf {\bibinfo {volume} {12}},\
  \bibinfo {pages} {974} (\bibinfo {year} {2021})}\BibitemShut {NoStop}%
\bibitem [{\citenamefont {Liu}\ \emph {et~al.}(2023{\natexlab{a}})\citenamefont
  {Liu}, \citenamefont {Ren}, \citenamefont {Huo}, \citenamefont {Cai},\ and\
  \citenamefont {Ning}}]{Liu2023}%
  \BibitemOpen
  \bibfield  {author} {\bibinfo {author} {\bibfnamefont {D.}~\bibnamefont
  {Liu}}, \bibinfo {author} {\bibfnamefont {Y.}~\bibnamefont {Ren}}, \bibinfo
  {author} {\bibfnamefont {Y.}~\bibnamefont {Huo}}, \bibinfo {author}
  {\bibfnamefont {Y.}~\bibnamefont {Cai}},\ and\ \bibinfo {author}
  {\bibfnamefont {T.}~\bibnamefont {Ning}},\ }\bibfield  {title} {\bibinfo
  {title} {Second harmonic generation in plasmonic metasurfaces enhanced by
  symmetry-protected dual bound states in the continuum},\ }\href
  {https://doi.org/10.1364/oe.496853} {\bibfield  {journal} {\bibinfo
  {journal} {Optics Express}\ }\textbf {\bibinfo {volume} {31}},\ \bibinfo
  {pages} {23127} (\bibinfo {year} {2023}{\natexlab{a}})}\BibitemShut {NoStop}%
\bibitem [{\citenamefont {Yang}\ \emph {et~al.}(2015)\citenamefont {Yang},
  \citenamefont {Wang}, \citenamefont {Boulesbaa}, \citenamefont {Kravchenko},
  \citenamefont {Briggs}, \citenamefont {Puretzky}, \citenamefont {Geohegan},\
  and\ \citenamefont {Valentine}}]{Yang2015}%
  \BibitemOpen
  \bibfield  {author} {\bibinfo {author} {\bibfnamefont {Y.}~\bibnamefont
  {Yang}}, \bibinfo {author} {\bibfnamefont {W.}~\bibnamefont {Wang}}, \bibinfo
  {author} {\bibfnamefont {A.}~\bibnamefont {Boulesbaa}}, \bibinfo {author}
  {\bibfnamefont {I.~I.}\ \bibnamefont {Kravchenko}}, \bibinfo {author}
  {\bibfnamefont {D.~P.}\ \bibnamefont {Briggs}}, \bibinfo {author}
  {\bibfnamefont {A.}~\bibnamefont {Puretzky}}, \bibinfo {author}
  {\bibfnamefont {D.}~\bibnamefont {Geohegan}},\ and\ \bibinfo {author}
  {\bibfnamefont {J.}~\bibnamefont {Valentine}},\ }\bibfield  {title} {\bibinfo
  {title} {Nonlinear fano-resonant dielectric metasurfaces},\ }\href
  {https://doi.org/10.1021/acs.nanolett.5b02802} {\bibfield  {journal}
  {\bibinfo  {journal} {Nano Letters}\ }\textbf {\bibinfo {volume} {15}},\
  \bibinfo {pages} {7388} (\bibinfo {year} {2015})}\BibitemShut {NoStop}%
\bibitem [{\citenamefont {Hu}\ \emph {et~al.}(2020)\citenamefont {Hu},
  \citenamefont {Wang}, \citenamefont {Luo}, \citenamefont {Ou}, \citenamefont
  {Li}, \citenamefont {Chen}, \citenamefont {Yang}, \citenamefont {Wang},\ and\
  \citenamefont {Duan}}]{Hu2020}%
  \BibitemOpen
  \bibfield  {author} {\bibinfo {author} {\bibfnamefont {Y.}~\bibnamefont
  {Hu}}, \bibinfo {author} {\bibfnamefont {X.}~\bibnamefont {Wang}}, \bibinfo
  {author} {\bibfnamefont {X.}~\bibnamefont {Luo}}, \bibinfo {author}
  {\bibfnamefont {X.}~\bibnamefont {Ou}}, \bibinfo {author} {\bibfnamefont
  {L.}~\bibnamefont {Li}}, \bibinfo {author} {\bibfnamefont {Y.}~\bibnamefont
  {Chen}}, \bibinfo {author} {\bibfnamefont {P.}~\bibnamefont {Yang}}, \bibinfo
  {author} {\bibfnamefont {S.}~\bibnamefont {Wang}},\ and\ \bibinfo {author}
  {\bibfnamefont {H.}~\bibnamefont {Duan}},\ }\bibfield  {title} {\bibinfo
  {title} {All-dielectric metasurfaces for polarization manipulation:
  principles and emerging applications},\ }\href
  {https://doi.org/10.1515/nanoph-2020-0220} {\bibfield  {journal} {\bibinfo
  {journal} {Nanophotonics}\ }\textbf {\bibinfo {volume} {9}},\ \bibinfo
  {pages} {3755} (\bibinfo {year} {2020})}\BibitemShut {NoStop}%
\bibitem [{\citenamefont {Alam}\ \emph {et~al.}(2018)\citenamefont {Alam},
  \citenamefont {Schulz}, \citenamefont {Upham}, \citenamefont {De~Leon},\ and\
  \citenamefont {Boyd}}]{Alam2018}%
  \BibitemOpen
  \bibfield  {author} {\bibinfo {author} {\bibfnamefont {M.~Z.}\ \bibnamefont
  {Alam}}, \bibinfo {author} {\bibfnamefont {S.~A.}\ \bibnamefont {Schulz}},
  \bibinfo {author} {\bibfnamefont {J.}~\bibnamefont {Upham}}, \bibinfo
  {author} {\bibfnamefont {I.}~\bibnamefont {De~Leon}},\ and\ \bibinfo {author}
  {\bibfnamefont {R.~W.}\ \bibnamefont {Boyd}},\ }\bibfield  {title} {\bibinfo
  {title} {Large optical nonlinearity of nanoantennas coupled to an
  epsilon-near-zero material},\ }\href
  {https://doi.org/10.1038/s41566-017-0089-9} {\bibfield  {journal} {\bibinfo
  {journal} {Nature Photonics}\ }\textbf {\bibinfo {volume} {12}},\ \bibinfo
  {pages} {79} (\bibinfo {year} {2018})}\BibitemShut {NoStop}%
\bibitem [{\citenamefont {Melik-Gaykazyan}\ \emph {et~al.}(2019)\citenamefont
  {Melik-Gaykazyan}, \citenamefont {Koshelev}, \citenamefont {Choi},
  \citenamefont {Kruk}, \citenamefont {Park}, \citenamefont {Fedyanin},\ and\
  \citenamefont {Kivshar}}]{MelikGaykazyan2019}%
  \BibitemOpen
  \bibfield  {author} {\bibinfo {author} {\bibfnamefont {E.~V.}\ \bibnamefont
  {Melik-Gaykazyan}}, \bibinfo {author} {\bibfnamefont {K.~L.}\ \bibnamefont
  {Koshelev}}, \bibinfo {author} {\bibfnamefont {J.-H.}\ \bibnamefont {Choi}},
  \bibinfo {author} {\bibfnamefont {S.}~\bibnamefont {Kruk}}, \bibinfo {author}
  {\bibfnamefont {H.-G.}\ \bibnamefont {Park}}, \bibinfo {author}
  {\bibfnamefont {A.~A.}\ \bibnamefont {Fedyanin}},\ and\ \bibinfo {author}
  {\bibfnamefont {Y.~S.}\ \bibnamefont {Kivshar}},\ }\bibfield  {title}
  {\bibinfo {title} {Enhanced second-harmonic generation with structured light
  in algaas nanoparticles governed by magnetic response},\ }\href
  {https://doi.org/10.1134/s0021364019020036} {\bibfield  {journal} {\bibinfo
  {journal} {JETP Letters}\ }\textbf {\bibinfo {volume} {109}},\ \bibinfo
  {pages} {131} (\bibinfo {year} {2019})}\BibitemShut {NoStop}%
\bibitem [{\citenamefont {Koshelev}\ \emph
  {et~al.}(2019{\natexlab{a}})\citenamefont {Koshelev}, \citenamefont {Tang},
  \citenamefont {Li}, \citenamefont {Choi}, \citenamefont {Li},\ and\
  \citenamefont {Kivshar}}]{Koshelev2019}%
  \BibitemOpen
  \bibfield  {author} {\bibinfo {author} {\bibfnamefont {K.}~\bibnamefont
  {Koshelev}}, \bibinfo {author} {\bibfnamefont {Y.}~\bibnamefont {Tang}},
  \bibinfo {author} {\bibfnamefont {K.}~\bibnamefont {Li}}, \bibinfo {author}
  {\bibfnamefont {D.-Y.}\ \bibnamefont {Choi}}, \bibinfo {author}
  {\bibfnamefont {G.}~\bibnamefont {Li}},\ and\ \bibinfo {author}
  {\bibfnamefont {Y.}~\bibnamefont {Kivshar}},\ }\bibfield  {title} {\bibinfo
  {title} {Nonlinear metasurfaces governed by bound states in the continuum},\
  }\href {https://doi.org/10.1021/acsphotonics.9b00700} {\bibfield  {journal}
  {\bibinfo  {journal} {ACS Photonics}\ }\textbf {\bibinfo {volume} {6}},\
  \bibinfo {pages} {1639} (\bibinfo {year} {2019}{\natexlab{a}})}\BibitemShut
  {NoStop}%
\bibitem [{\citenamefont {Qi}\ \emph {et~al.}(2023)\citenamefont {Qi},
  \citenamefont {Wu}, \citenamefont {Wu}, \citenamefont {Ren}, \citenamefont
  {Wei}, \citenamefont {Wang}, \citenamefont {Jiang}, \citenamefont {Li},
  \citenamefont {Guo}, \citenamefont {Yang}, \citenamefont {Zheng},
  \citenamefont {Sun},\ and\ \citenamefont {Chen}}]{Qi2023}%
  \BibitemOpen
  \bibfield  {author} {\bibinfo {author} {\bibfnamefont {X.}~\bibnamefont
  {Qi}}, \bibinfo {author} {\bibfnamefont {J.}~\bibnamefont {Wu}}, \bibinfo
  {author} {\bibfnamefont {F.}~\bibnamefont {Wu}}, \bibinfo {author}
  {\bibfnamefont {M.}~\bibnamefont {Ren}}, \bibinfo {author} {\bibfnamefont
  {Q.}~\bibnamefont {Wei}}, \bibinfo {author} {\bibfnamefont {Y.}~\bibnamefont
  {Wang}}, \bibinfo {author} {\bibfnamefont {H.}~\bibnamefont {Jiang}},
  \bibinfo {author} {\bibfnamefont {Y.}~\bibnamefont {Li}}, \bibinfo {author}
  {\bibfnamefont {Z.}~\bibnamefont {Guo}}, \bibinfo {author} {\bibfnamefont
  {Y.}~\bibnamefont {Yang}}, \bibinfo {author} {\bibfnamefont {W.}~\bibnamefont
  {Zheng}}, \bibinfo {author} {\bibfnamefont {Y.}~\bibnamefont {Sun}},\ and\
  \bibinfo {author} {\bibfnamefont {H.}~\bibnamefont {Chen}},\ }\bibfield
  {title} {\bibinfo {title} {Steerable merging bound states in the continuum on
  a quasi-flatband of photonic crystal slabs without breaking symmetry},\
  }\href {https://doi.org/10.1364/prj.487665} {\bibfield  {journal} {\bibinfo
  {journal} {Photonics Research}\ }\textbf {\bibinfo {volume} {11}},\ \bibinfo
  {pages} {1262} (\bibinfo {year} {2023})}\BibitemShut {NoStop}%
\bibitem [{\citenamefont {Sun}\ \emph {et~al.}(2024{\natexlab{a}})\citenamefont
  {Sun}, \citenamefont {Wang}, \citenamefont {Xie},\ and\ \citenamefont
  {Zhao}}]{Sun2024}%
  \BibitemOpen
  \bibfield  {author} {\bibinfo {author} {\bibfnamefont {G.}~\bibnamefont
  {Sun}}, \bibinfo {author} {\bibfnamefont {Y.}~\bibnamefont {Wang}}, \bibinfo
  {author} {\bibfnamefont {R.}~\bibnamefont {Xie}},\ and\ \bibinfo {author}
  {\bibfnamefont {X.}~\bibnamefont {Zhao}},\ }\bibfield  {title} {\bibinfo
  {title} {Polarization-insensitive terahertz third-harmonic generation from
  degenerate pairs of mirror-coupled super-bics},\ }\href
  {https://doi.org/10.1063/5.0221133} {\bibfield  {journal} {\bibinfo
  {journal} {Applied Physics Letters}\ }\textbf {\bibinfo {volume} {125}},\
  \bibinfo {pages} {081702} (\bibinfo {year} {2024}{\natexlab{a}})}\BibitemShut
  {NoStop}%
\bibitem [{\citenamefont {Tu}\ \emph {et~al.}(2024)\citenamefont {Tu},
  \citenamefont {Feng}, \citenamefont {Li}, \citenamefont {Xing}, \citenamefont
  {Wu}, \citenamefont {Liu},\ and\ \citenamefont {Xiao}}]{Tu2024}%
  \BibitemOpen
  \bibfield  {author} {\bibinfo {author} {\bibfnamefont {X.}~\bibnamefont
  {Tu}}, \bibinfo {author} {\bibfnamefont {S.}~\bibnamefont {Feng}}, \bibinfo
  {author} {\bibfnamefont {J.}~\bibnamefont {Li}}, \bibinfo {author}
  {\bibfnamefont {Y.}~\bibnamefont {Xing}}, \bibinfo {author} {\bibfnamefont
  {F.}~\bibnamefont {Wu}}, \bibinfo {author} {\bibfnamefont {T.}~\bibnamefont
  {Liu}},\ and\ \bibinfo {author} {\bibfnamefont {S.}~\bibnamefont {Xiao}},\
  }\bibfield  {title} {\bibinfo {title} {Enhanced second-harmonic generation in
  high-q all-dielectric metasurfaces with backward frequency conversion},\
  }\href {https://doi.org/10.1103/physreva.109.063522} {\bibfield  {journal}
  {\bibinfo  {journal} {Physical Review A}\ }\textbf {\bibinfo {volume}
  {109}},\ \bibinfo {pages} {063522} (\bibinfo {year} {2024})}\BibitemShut
  {NoStop}%
\bibitem [{\citenamefont {Hsu}\ \emph {et~al.}(2016)\citenamefont {Hsu},
  \citenamefont {Zhen}, \citenamefont {Stone}, \citenamefont {Joannopoulos},\
  and\ \citenamefont {Solja{\v{c}}i{\'c}}}]{hsu2016bound}%
  \BibitemOpen
  \bibfield  {author} {\bibinfo {author} {\bibfnamefont {C.~W.}\ \bibnamefont
  {Hsu}}, \bibinfo {author} {\bibfnamefont {B.}~\bibnamefont {Zhen}}, \bibinfo
  {author} {\bibfnamefont {A.~D.}\ \bibnamefont {Stone}}, \bibinfo {author}
  {\bibfnamefont {J.~D.}\ \bibnamefont {Joannopoulos}},\ and\ \bibinfo {author}
  {\bibfnamefont {M.}~\bibnamefont {Solja{\v{c}}i{\'c}}},\ }\bibfield  {title}
  {\bibinfo {title} {Bound states in the continuum},\ }\href@noop {} {\bibfield
   {journal} {\bibinfo  {journal} {Nature Reviews Materials}\ }\textbf
  {\bibinfo {volume} {1}},\ \bibinfo {pages} {16048} (\bibinfo {year}
  {2016})}\BibitemShut {NoStop}%
\bibitem [{\citenamefont {Friedrich}\ and\ \citenamefont
  {Wintgen}(1985)}]{friedrich1985interfering}%
  \BibitemOpen
  \bibfield  {author} {\bibinfo {author} {\bibfnamefont {H.}~\bibnamefont
  {Friedrich}}\ and\ \bibinfo {author} {\bibfnamefont {D.}~\bibnamefont
  {Wintgen}},\ }\bibfield  {title} {\bibinfo {title} {Interfering resonances
  and bound states in the continuum},\ }\href@noop {} {\bibfield  {journal}
  {\bibinfo  {journal} {Physical Review A}\ }\textbf {\bibinfo {volume} {32}},\
  \bibinfo {pages} {3231} (\bibinfo {year} {1985})}\BibitemShut {NoStop}%
\bibitem [{\citenamefont {Koshelev}\ \emph
  {et~al.}(2019{\natexlab{b}})\citenamefont {Koshelev}, \citenamefont
  {Favraud}, \citenamefont {Bogdanov}, \citenamefont {Kivshar},\ and\
  \citenamefont {Fratalocchi}}]{koshelev2019nonradiating}%
  \BibitemOpen
  \bibfield  {author} {\bibinfo {author} {\bibfnamefont {K.}~\bibnamefont
  {Koshelev}}, \bibinfo {author} {\bibfnamefont {G.}~\bibnamefont {Favraud}},
  \bibinfo {author} {\bibfnamefont {A.}~\bibnamefont {Bogdanov}}, \bibinfo
  {author} {\bibfnamefont {Y.}~\bibnamefont {Kivshar}},\ and\ \bibinfo {author}
  {\bibfnamefont {A.}~\bibnamefont {Fratalocchi}},\ }\bibfield  {title}
  {\bibinfo {title} {Nonradiating photonics with resonant dielectric
  nanostructures},\ }\href@noop {} {\bibfield  {journal} {\bibinfo  {journal}
  {Nanophotonics}\ }\textbf {\bibinfo {volume} {8}},\ \bibinfo {pages} {725}
  (\bibinfo {year} {2019}{\natexlab{b}})}\BibitemShut {NoStop}%
\bibitem [{\citenamefont {Wang}\ \emph {et~al.}(2020)\citenamefont {Wang},
  \citenamefont {Duan}, \citenamefont {Chen}, \citenamefont {Zhou},
  \citenamefont {Liu},\ and\ \citenamefont {Xiao}}]{Wang2020}%
  \BibitemOpen
  \bibfield  {author} {\bibinfo {author} {\bibfnamefont {X.}~\bibnamefont
  {Wang}}, \bibinfo {author} {\bibfnamefont {J.}~\bibnamefont {Duan}}, \bibinfo
  {author} {\bibfnamefont {W.}~\bibnamefont {Chen}}, \bibinfo {author}
  {\bibfnamefont {C.}~\bibnamefont {Zhou}}, \bibinfo {author} {\bibfnamefont
  {T.}~\bibnamefont {Liu}},\ and\ \bibinfo {author} {\bibfnamefont
  {S.}~\bibnamefont {Xiao}},\ }\bibfield  {title} {\bibinfo {title}
  {Controlling light absorption of graphene at critical coupling through
  magnetic dipole quasi-bound states in the continuum resonance},\ }\href
  {https://doi.org/10.1103/physrevb.102.155432} {\bibfield  {journal} {\bibinfo
   {journal} {Physical Review B}\ }\textbf {\bibinfo {volume} {102}},\ \bibinfo
  {pages} {155432} (\bibinfo {year} {2020})}\BibitemShut {NoStop}%
\bibitem [{\citenamefont {Wu}\ \emph {et~al.}(2021)\citenamefont {Wu},
  \citenamefont {Luo}, \citenamefont {Wu}, \citenamefont {Fan}, \citenamefont
  {Qi}, \citenamefont {Jian}, \citenamefont {Liu}, \citenamefont {Xiao},
  \citenamefont {Chen}, \citenamefont {Jiang}, \citenamefont {Sun},\ and\
  \citenamefont {Chen}}]{Wu2021}%
  \BibitemOpen
  \bibfield  {author} {\bibinfo {author} {\bibfnamefont {F.}~\bibnamefont
  {Wu}}, \bibinfo {author} {\bibfnamefont {M.}~\bibnamefont {Luo}}, \bibinfo
  {author} {\bibfnamefont {J.}~\bibnamefont {Wu}}, \bibinfo {author}
  {\bibfnamefont {C.}~\bibnamefont {Fan}}, \bibinfo {author} {\bibfnamefont
  {X.}~\bibnamefont {Qi}}, \bibinfo {author} {\bibfnamefont {Y.}~\bibnamefont
  {Jian}}, \bibinfo {author} {\bibfnamefont {D.}~\bibnamefont {Liu}}, \bibinfo
  {author} {\bibfnamefont {S.}~\bibnamefont {Xiao}}, \bibinfo {author}
  {\bibfnamefont {G.}~\bibnamefont {Chen}}, \bibinfo {author} {\bibfnamefont
  {H.}~\bibnamefont {Jiang}}, \bibinfo {author} {\bibfnamefont
  {Y.}~\bibnamefont {Sun}},\ and\ \bibinfo {author} {\bibfnamefont
  {H.}~\bibnamefont {Chen}},\ }\bibfield  {title} {\bibinfo {title} {Dual
  quasibound states in the continuum in compound grating waveguide structures
  for large positive and negative goos-hänchen shifts with perfect
  reflection},\ }\href {https://doi.org/10.1103/physreva.104.023518} {\bibfield
   {journal} {\bibinfo  {journal} {Physical Review A}\ }\textbf {\bibinfo
  {volume} {104}},\ \bibinfo {pages} {023518} (\bibinfo {year}
  {2021})}\BibitemShut {NoStop}%
\bibitem [{\citenamefont {Qin}\ \emph {et~al.}(2021)\citenamefont {Qin},
  \citenamefont {Xiao}, \citenamefont {Liu}, \citenamefont {Ouyang},
  \citenamefont {Yu}, \citenamefont {Wang},\ and\ \citenamefont
  {Liao}}]{Qin2021}%
  \BibitemOpen
  \bibfield  {author} {\bibinfo {author} {\bibfnamefont {M.}~\bibnamefont
  {Qin}}, \bibinfo {author} {\bibfnamefont {S.}~\bibnamefont {Xiao}}, \bibinfo
  {author} {\bibfnamefont {W.}~\bibnamefont {Liu}}, \bibinfo {author}
  {\bibfnamefont {M.}~\bibnamefont {Ouyang}}, \bibinfo {author} {\bibfnamefont
  {T.}~\bibnamefont {Yu}}, \bibinfo {author} {\bibfnamefont {T.}~\bibnamefont
  {Wang}},\ and\ \bibinfo {author} {\bibfnamefont {Q.}~\bibnamefont {Liao}},\
  }\bibfield  {title} {\bibinfo {title} {Strong coupling between excitons and
  magnetic dipole quasi-bound states in the continuum in ws2-tio2 hybrid
  metasurfaces},\ }\href {https://doi.org/10.1364/oe.427141} {\bibfield
  {journal} {\bibinfo  {journal} {Optics Express}\ }\textbf {\bibinfo {volume}
  {29}},\ \bibinfo {pages} {18026} (\bibinfo {year} {2021})}\BibitemShut
  {NoStop}%
\bibitem [{\citenamefont {Xie}\ \emph {et~al.}(2021)\citenamefont {Xie},
  \citenamefont {Liang}, \citenamefont {Jia}, \citenamefont {Li}, \citenamefont
  {Chen}, \citenamefont {Chang}, \citenamefont {Zhang},\ and\ \citenamefont
  {Wang}}]{Xie2021}%
  \BibitemOpen
  \bibfield  {author} {\bibinfo {author} {\bibfnamefont {P.}~\bibnamefont
  {Xie}}, \bibinfo {author} {\bibfnamefont {Z.}~\bibnamefont {Liang}}, \bibinfo
  {author} {\bibfnamefont {T.}~\bibnamefont {Jia}}, \bibinfo {author}
  {\bibfnamefont {D.}~\bibnamefont {Li}}, \bibinfo {author} {\bibfnamefont
  {Y.}~\bibnamefont {Chen}}, \bibinfo {author} {\bibfnamefont {P.}~\bibnamefont
  {Chang}}, \bibinfo {author} {\bibfnamefont {H.}~\bibnamefont {Zhang}},\ and\
  \bibinfo {author} {\bibfnamefont {W.}~\bibnamefont {Wang}},\ }\bibfield
  {title} {\bibinfo {title} {Strong coupling between excitons in a
  two-dimensional atomic crystal and quasibound states in the continuum in a
  two-dimensional all-dielectric asymmetric metasurface},\ }\href
  {https://doi.org/10.1103/physrevb.104.125446} {\bibfield  {journal} {\bibinfo
   {journal} {Physical Review B}\ }\textbf {\bibinfo {volume} {104}},\ \bibinfo
  {pages} {125446} (\bibinfo {year} {2021})}\BibitemShut {NoStop}%
\bibitem [{\citenamefont {Zeng}\ \emph {et~al.}(2021)\citenamefont {Zeng},
  \citenamefont {Liu}, \citenamefont {Wang},\ and\ \citenamefont
  {Lin}}]{Zeng2021}%
  \BibitemOpen
  \bibfield  {author} {\bibinfo {author} {\bibfnamefont {T.-Y.}\ \bibnamefont
  {Zeng}}, \bibinfo {author} {\bibfnamefont {G.-D.}\ \bibnamefont {Liu}},
  \bibinfo {author} {\bibfnamefont {L.-L.}\ \bibnamefont {Wang}},\ and\
  \bibinfo {author} {\bibfnamefont {Q.}~\bibnamefont {Lin}},\ }\bibfield
  {title} {\bibinfo {title} {Light-matter interactions enhanced by quasi-bound
  states in the continuum in a graphene-dielectric metasurface},\ }\href
  {https://doi.org/10.1364/oe.446072} {\bibfield  {journal} {\bibinfo
  {journal} {Optics Express}\ }\textbf {\bibinfo {volume} {29}},\ \bibinfo
  {pages} {40177} (\bibinfo {year} {2021})}\BibitemShut {NoStop}%
\bibitem [{\citenamefont {Qu}\ \emph {et~al.}(2022)\citenamefont {Qu},
  \citenamefont {Bai}, \citenamefont {Jin}, \citenamefont {Liu}, \citenamefont
  {Wu}, \citenamefont {Gao}, \citenamefont {Li}, \citenamefont {Cai},
  \citenamefont {Ren},\ and\ \citenamefont {Xu}}]{qu2022giant}%
  \BibitemOpen
  \bibfield  {author} {\bibinfo {author} {\bibfnamefont {L.}~\bibnamefont
  {Qu}}, \bibinfo {author} {\bibfnamefont {L.}~\bibnamefont {Bai}}, \bibinfo
  {author} {\bibfnamefont {C.}~\bibnamefont {Jin}}, \bibinfo {author}
  {\bibfnamefont {Q.}~\bibnamefont {Liu}}, \bibinfo {author} {\bibfnamefont
  {W.}~\bibnamefont {Wu}}, \bibinfo {author} {\bibfnamefont {B.}~\bibnamefont
  {Gao}}, \bibinfo {author} {\bibfnamefont {J.}~\bibnamefont {Li}}, \bibinfo
  {author} {\bibfnamefont {W.}~\bibnamefont {Cai}}, \bibinfo {author}
  {\bibfnamefont {M.}~\bibnamefont {Ren}},\ and\ \bibinfo {author}
  {\bibfnamefont {J.}~\bibnamefont {Xu}},\ }\bibfield  {title} {\bibinfo
  {title} {Giant second harmonic generation from membrane metasurfaces},\
  }\href@noop {} {\bibfield  {journal} {\bibinfo  {journal} {Nano Letters}\
  }\textbf {\bibinfo {volume} {22}},\ \bibinfo {pages} {9652} (\bibinfo {year}
  {2022})}\BibitemShut {NoStop}%
\bibitem [{\citenamefont {Liu}\ \emph {et~al.}(2023{\natexlab{b}})\citenamefont
  {Liu}, \citenamefont {Qu}, \citenamefont {Gu}, \citenamefont {Zhang},
  \citenamefont {Wu}, \citenamefont {Cai}, \citenamefont {Ren},\ and\
  \citenamefont {Xu}}]{liu2023boosting}%
  \BibitemOpen
  \bibfield  {author} {\bibinfo {author} {\bibfnamefont {Q.}~\bibnamefont
  {Liu}}, \bibinfo {author} {\bibfnamefont {L.}~\bibnamefont {Qu}}, \bibinfo
  {author} {\bibfnamefont {Z.}~\bibnamefont {Gu}}, \bibinfo {author}
  {\bibfnamefont {D.}~\bibnamefont {Zhang}}, \bibinfo {author} {\bibfnamefont
  {W.}~\bibnamefont {Wu}}, \bibinfo {author} {\bibfnamefont {W.}~\bibnamefont
  {Cai}}, \bibinfo {author} {\bibfnamefont {M.}~\bibnamefont {Ren}},\ and\
  \bibinfo {author} {\bibfnamefont {J.}~\bibnamefont {Xu}},\ }\bibfield
  {title} {\bibinfo {title} {Boosting second harmonic generation by merging
  bound states in the continuum},\ }\href@noop {} {\bibfield  {journal}
  {\bibinfo  {journal} {Physical Review B}\ }\textbf {\bibinfo {volume}
  {107}},\ \bibinfo {pages} {245408} (\bibinfo {year}
  {2023}{\natexlab{b}})}\BibitemShut {NoStop}%
\bibitem [{\citenamefont {Huang}\ \emph {et~al.}(2023)\citenamefont {Huang},
  \citenamefont {Xu}, \citenamefont {Powell}, \citenamefont {Padilla},\ and\
  \citenamefont {Miroshnichenko}}]{huang2023resonant}%
  \BibitemOpen
  \bibfield  {author} {\bibinfo {author} {\bibfnamefont {L.}~\bibnamefont
  {Huang}}, \bibinfo {author} {\bibfnamefont {L.}~\bibnamefont {Xu}}, \bibinfo
  {author} {\bibfnamefont {D.~A.}\ \bibnamefont {Powell}}, \bibinfo {author}
  {\bibfnamefont {W.~J.}\ \bibnamefont {Padilla}},\ and\ \bibinfo {author}
  {\bibfnamefont {A.~E.}\ \bibnamefont {Miroshnichenko}},\ }\bibfield  {title}
  {\bibinfo {title} {Resonant leaky modes in all-dielectric metasystems:
  Fundamentals and applications},\ }\href@noop {} {\bibfield  {journal}
  {\bibinfo  {journal} {Physics Reports}\ }\textbf {\bibinfo {volume} {1008}},\
  \bibinfo {pages} {1} (\bibinfo {year} {2023})}\BibitemShut {NoStop}%
\bibitem [{\citenamefont {Liu}\ \emph {et~al.}(2024{\natexlab{a}})\citenamefont
  {Liu}, \citenamefont {Qiu}, \citenamefont {Xu}, \citenamefont {Qin},
  \citenamefont {Wan}, \citenamefont {Yu}, \citenamefont {Liu}, \citenamefont
  {Huang},\ and\ \citenamefont {Xiao}}]{Liu2024}%
  \BibitemOpen
  \bibfield  {author} {\bibinfo {author} {\bibfnamefont {T.}~\bibnamefont
  {Liu}}, \bibinfo {author} {\bibfnamefont {J.}~\bibnamefont {Qiu}}, \bibinfo
  {author} {\bibfnamefont {L.}~\bibnamefont {Xu}}, \bibinfo {author}
  {\bibfnamefont {M.}~\bibnamefont {Qin}}, \bibinfo {author} {\bibfnamefont
  {L.}~\bibnamefont {Wan}}, \bibinfo {author} {\bibfnamefont {T.}~\bibnamefont
  {Yu}}, \bibinfo {author} {\bibfnamefont {Q.}~\bibnamefont {Liu}}, \bibinfo
  {author} {\bibfnamefont {L.}~\bibnamefont {Huang}},\ and\ \bibinfo {author}
  {\bibfnamefont {S.}~\bibnamefont {Xiao}},\ }\bibfield  {title} {\bibinfo
  {title} {Edge detection imaging by quasi-bound states in the continuum},\
  }\href {https://doi.org/10.1021/acs.nanolett.4c04543} {\bibfield  {journal}
  {\bibinfo  {journal} {Nano Letters}\ }\textbf {\bibinfo {volume} {24}},\
  \bibinfo {pages} {14466} (\bibinfo {year} {2024}{\natexlab{a}})}\BibitemShut
  {NoStop}%
\bibitem [{\citenamefont {Zhang}\ \emph {et~al.}(2024)\citenamefont {Zhang},
  \citenamefont {Liu}, \citenamefont {Lei}, \citenamefont {Deng}, \citenamefont
  {Wang}, \citenamefont {Liao}, \citenamefont {Liu}, \citenamefont {Xiao},\
  and\ \citenamefont {Yu}}]{Zhang2024}%
  \BibitemOpen
  \bibfield  {author} {\bibinfo {author} {\bibfnamefont {D.}~\bibnamefont
  {Zhang}}, \bibinfo {author} {\bibfnamefont {T.}~\bibnamefont {Liu}}, \bibinfo
  {author} {\bibfnamefont {L.}~\bibnamefont {Lei}}, \bibinfo {author}
  {\bibfnamefont {W.}~\bibnamefont {Deng}}, \bibinfo {author} {\bibfnamefont
  {T.}~\bibnamefont {Wang}}, \bibinfo {author} {\bibfnamefont {Q.}~\bibnamefont
  {Liao}}, \bibinfo {author} {\bibfnamefont {W.}~\bibnamefont {Liu}}, \bibinfo
  {author} {\bibfnamefont {S.}~\bibnamefont {Xiao}},\ and\ \bibinfo {author}
  {\bibfnamefont {T.}~\bibnamefont {Yu}},\ }\bibfield  {title} {\bibinfo
  {title} {Tailoring intrinsic chirality in a two-dimensional planar waveguide
  grating via quasibound states in the continuum},\ }\href
  {https://doi.org/10.1103/physrevb.109.205403} {\bibfield  {journal} {\bibinfo
   {journal} {Physical Review B}\ }\textbf {\bibinfo {volume} {109}},\ \bibinfo
  {pages} {205403} (\bibinfo {year} {2024})}\BibitemShut {NoStop}%
\bibitem [{\citenamefont {Liu}\ \emph {et~al.}(2024{\natexlab{b}})\citenamefont
  {Liu}, \citenamefont {Qin}, \citenamefont {Feng}, \citenamefont {Tu},
  \citenamefont {Guo}, \citenamefont {Wu},\ and\ \citenamefont
  {Xiao}}]{Liu2024a}%
  \BibitemOpen
  \bibfield  {author} {\bibinfo {author} {\bibfnamefont {T.}~\bibnamefont
  {Liu}}, \bibinfo {author} {\bibfnamefont {M.}~\bibnamefont {Qin}}, \bibinfo
  {author} {\bibfnamefont {S.}~\bibnamefont {Feng}}, \bibinfo {author}
  {\bibfnamefont {X.}~\bibnamefont {Tu}}, \bibinfo {author} {\bibfnamefont
  {T.}~\bibnamefont {Guo}}, \bibinfo {author} {\bibfnamefont {F.}~\bibnamefont
  {Wu}},\ and\ \bibinfo {author} {\bibfnamefont {S.}~\bibnamefont {Xiao}},\
  }\bibfield  {title} {\bibinfo {title} {Efficient photon-pair generation
  empowered by dual quasibound states in the continuum},\ }\href
  {https://doi.org/10.1103/phys revb.109.155424} {\bibfield  {journal}
  {\bibinfo  {journal} {Physical Review B}\ }\textbf {\bibinfo {volume}
  {109}},\ \bibinfo {pages} {155424} (\bibinfo {year}
  {2024}{\natexlab{b}})}\BibitemShut {NoStop}%
\bibitem [{\citenamefont {Hajian}\ \emph {et~al.}(2024)\citenamefont {Hajian},
  \citenamefont {Zhang}, \citenamefont {McCormack}, \citenamefont {Zhang},
  \citenamefont {Dobie}, \citenamefont {Rukhlenko}, \citenamefont {Ozbay},\
  and\ \citenamefont {Louise~Bradley}}]{hajian2024quasi}%
  \BibitemOpen
  \bibfield  {author} {\bibinfo {author} {\bibfnamefont {H.}~\bibnamefont
  {Hajian}}, \bibinfo {author} {\bibfnamefont {X.}~\bibnamefont {Zhang}},
  \bibinfo {author} {\bibfnamefont {O.}~\bibnamefont {McCormack}}, \bibinfo
  {author} {\bibfnamefont {Y.}~\bibnamefont {Zhang}}, \bibinfo {author}
  {\bibfnamefont {J.}~\bibnamefont {Dobie}}, \bibinfo {author} {\bibfnamefont
  {I.~D.}\ \bibnamefont {Rukhlenko}}, \bibinfo {author} {\bibfnamefont
  {E.}~\bibnamefont {Ozbay}},\ and\ \bibinfo {author} {\bibfnamefont
  {A.}~\bibnamefont {Louise~Bradley}},\ }\bibfield  {title} {\bibinfo {title}
  {Quasi-bound states in the continuum for electro magnetic induced
  transparency and strong excitonic coupling},\ }\href@noop {} {\bibfield
  {journal} {\bibinfo  {journal} {Optics Express}\ }\textbf {\bibinfo {volume}
  {32}},\ \bibinfo {pages} {19163} (\bibinfo {year} {2024})}\BibitemShut
  {NoStop}%
\bibitem [{\citenamefont {Jiang}\ \emph {et~al.}(2024)\citenamefont {Jiang},
  \citenamefont {Sun}, \citenamefont {Jia}, \citenamefont {Cai}, \citenamefont
  {Levy},\ and\ \citenamefont {Han}}]{jiang2024tunable}%
  \BibitemOpen
  \bibfield  {author} {\bibinfo {author} {\bibfnamefont {H.}~\bibnamefont
  {Jiang}}, \bibinfo {author} {\bibfnamefont {K.}~\bibnamefont {Sun}}, \bibinfo
  {author} {\bibfnamefont {Y.}~\bibnamefont {Jia}}, \bibinfo {author}
  {\bibfnamefont {Y.}~\bibnamefont {Cai}}, \bibinfo {author} {\bibfnamefont
  {U.}~\bibnamefont {Levy}},\ and\ \bibinfo {author} {\bibfnamefont
  {Z.}~\bibnamefont {Han}},\ }\bibfield  {title} {\bibinfo {title} {Tunable
  second harmonic generation with large enhancement in a nonlocal
  all-dielectric metasurface over a broad spectral range},\ }\href@noop {}
  {\bibfield  {journal} {\bibinfo  {journal} {Advanced Optical Materials}\ ,\
  \bibinfo {pages} {2303229}} (\bibinfo {year} {2024})}\BibitemShut {NoStop}%
\bibitem [{\citenamefont {Carletti}\ \emph {et~al.}(2019)\citenamefont
  {Carletti}, \citenamefont {Kruk}, \citenamefont {Bogdanov}, \citenamefont
  {De~Angelis},\ and\ \citenamefont {Kivshar}}]{Carletti2019}%
  \BibitemOpen
  \bibfield  {author} {\bibinfo {author} {\bibfnamefont {L.}~\bibnamefont
  {Carletti}}, \bibinfo {author} {\bibfnamefont {S.~S.}\ \bibnamefont {Kruk}},
  \bibinfo {author} {\bibfnamefont {A.~A.}\ \bibnamefont {Bogdanov}}, \bibinfo
  {author} {\bibfnamefont {C.}~\bibnamefont {De~Angelis}},\ and\ \bibinfo
  {author} {\bibfnamefont {Y.}~\bibnamefont {Kivshar}},\ }\bibfield  {title}
  {\bibinfo {title} {High-harmonic generation at the nanoscale boosted by bound
  states in the continuum},\ }\href
  {https://doi.org/10.1103/physrevresearch.1.023016} {\bibfield  {journal}
  {\bibinfo  {journal} {Physical Review Research}\ }\textbf {\bibinfo {volume}
  {1}},\ \bibinfo {pages} {023016} (\bibinfo {year} {2019})}\BibitemShut
  {NoStop}%
\bibitem [{\citenamefont {Ning}\ \emph {et~al.}(2020)\citenamefont {Ning},
  \citenamefont {Li}, \citenamefont {Zhao}, \citenamefont {Yin}, \citenamefont
  {Huo}, \citenamefont {Zhao},\ and\ \citenamefont {Yue}}]{Ning2020}%
  \BibitemOpen
  \bibfield  {author} {\bibinfo {author} {\bibfnamefont {T.}~\bibnamefont
  {Ning}}, \bibinfo {author} {\bibfnamefont {X.}~\bibnamefont {Li}}, \bibinfo
  {author} {\bibfnamefont {Y.}~\bibnamefont {Zhao}}, \bibinfo {author}
  {\bibfnamefont {L.}~\bibnamefont {Yin}}, \bibinfo {author} {\bibfnamefont
  {Y.}~\bibnamefont {Huo}}, \bibinfo {author} {\bibfnamefont {L.}~\bibnamefont
  {Zhao}},\ and\ \bibinfo {author} {\bibfnamefont {Q.}~\bibnamefont {Yue}},\
  }\bibfield  {title} {\bibinfo {title} {Giant enhancement of harmonic
  generation in all-dielectric resonant waveguide gratings of quasi-bound
  states in the continuum},\ }\href {https://doi.org/10.1364/oe.409276}
  {\bibfield  {journal} {\bibinfo  {journal} {Optics Express}\ }\textbf
  {\bibinfo {volume} {28}},\ \bibinfo {pages} {34024} (\bibinfo {year}
  {2020})}\BibitemShut {NoStop}%
\bibitem [{\citenamefont {Xiao}\ \emph {et~al.}(2022)\citenamefont {Xiao},
  \citenamefont {Qin}, \citenamefont {Duan}, \citenamefont {Wu},\ and\
  \citenamefont {Liu}}]{Xiao2022}%
  \BibitemOpen
  \bibfield  {author} {\bibinfo {author} {\bibfnamefont {S.}~\bibnamefont
  {Xiao}}, \bibinfo {author} {\bibfnamefont {M.}~\bibnamefont {Qin}}, \bibinfo
  {author} {\bibfnamefont {J.}~\bibnamefont {Duan}}, \bibinfo {author}
  {\bibfnamefont {F.}~\bibnamefont {Wu}},\ and\ \bibinfo {author}
  {\bibfnamefont {T.}~\bibnamefont {Liu}},\ }\bibfield  {title} {\bibinfo
  {title} {Polarization-controlled dynamically switchable high-harmonic
  generation from all-dielectric metasurfaces governed by dual bound states in
  the continuum},\ }\href {https://doi.org/10.1103/physrevb.105.195440}
  {\bibfield  {journal} {\bibinfo  {journal} {Physical Review B}\ }\textbf
  {\bibinfo {volume} {105}},\ \bibinfo {pages} {195440} (\bibinfo {year}
  {2022})}\BibitemShut {NoStop}%
\bibitem [{\citenamefont {Liu}\ \emph {et~al.}(2023{\natexlab{c}})\citenamefont
  {Liu}, \citenamefont {Qin}, \citenamefont {Wu},\ and\ \citenamefont
  {Xiao}}]{Liu2023a}%
  \BibitemOpen
  \bibfield  {author} {\bibinfo {author} {\bibfnamefont {T.}~\bibnamefont
  {Liu}}, \bibinfo {author} {\bibfnamefont {M.}~\bibnamefont {Qin}}, \bibinfo
  {author} {\bibfnamefont {F.}~\bibnamefont {Wu}},\ and\ \bibinfo {author}
  {\bibfnamefont {S.}~\bibnamefont {Xiao}},\ }\bibfield  {title} {\bibinfo
  {title} {High-efficiency optical frequency mixing in an all-dielectric
  metasurface enabled by multiple bound states in the continuum},\ }\href
  {https://doi.org/10.1103/physrevb.107.075441} {\bibfield  {journal} {\bibinfo
   {journal} {Physical Review B}\ }\textbf {\bibinfo {volume} {107}},\ \bibinfo
  {pages} {075441} (\bibinfo {year} {2023}{\natexlab{c}})}\BibitemShut
  {NoStop}%
\bibitem [{\citenamefont {Tang}\ \emph {et~al.}(2024)\citenamefont {Tang},
  \citenamefont {Zhao}, \citenamefont {Wang}, \citenamefont {Gao},
  \citenamefont {He}, \citenamefont {Zhu}, \citenamefont {Wang}, \citenamefont
  {Yu}, \citenamefont {Peng},\ and\ \citenamefont {Wang}}]{Tang2024}%
  \BibitemOpen
  \bibfield  {author} {\bibinfo {author} {\bibfnamefont {W.}~\bibnamefont
  {Tang}}, \bibinfo {author} {\bibfnamefont {Q.}~\bibnamefont {Zhao}}, \bibinfo
  {author} {\bibfnamefont {Z.}~\bibnamefont {Wang}}, \bibinfo {author}
  {\bibfnamefont {Y.}~\bibnamefont {Gao}}, \bibinfo {author} {\bibfnamefont
  {J.}~\bibnamefont {He}}, \bibinfo {author} {\bibfnamefont {Y.}~\bibnamefont
  {Zhu}}, \bibinfo {author} {\bibfnamefont {S.}~\bibnamefont {Wang}}, \bibinfo
  {author} {\bibfnamefont {H.}~\bibnamefont {Yu}}, \bibinfo {author}
  {\bibfnamefont {R.}~\bibnamefont {Peng}},\ and\ \bibinfo {author}
  {\bibfnamefont {M.}~\bibnamefont {Wang}},\ }\bibfield  {title} {\bibinfo
  {title} {Realizing high-efficiency third harmonic generation via accidental
  bound states in the continuum},\ }\href {https://doi.org/10.1364/ol.514828}
  {\bibfield  {journal} {\bibinfo  {journal} {Optics Letters}\ }\textbf
  {\bibinfo {volume} {49}},\ \bibinfo {pages} {1169} (\bibinfo {year}
  {2024})}\BibitemShut {NoStop}%
\bibitem [{\citenamefont {Qin}\ \emph {et~al.}(2024)\citenamefont {Qin},
  \citenamefont {Wei}, \citenamefont {Xu}, \citenamefont {Ma}, \citenamefont
  {Li}, \citenamefont {Gao}, \citenamefont {Liu},\ and\ \citenamefont
  {Wu}}]{Qin2024}%
  \BibitemOpen
  \bibfield  {author} {\bibinfo {author} {\bibfnamefont {M.}~\bibnamefont
  {Qin}}, \bibinfo {author} {\bibfnamefont {G.}~\bibnamefont {Wei}}, \bibinfo
  {author} {\bibfnamefont {H.}~\bibnamefont {Xu}}, \bibinfo {author}
  {\bibfnamefont {R.}~\bibnamefont {Ma}}, \bibinfo {author} {\bibfnamefont
  {H.}~\bibnamefont {Li}}, \bibinfo {author} {\bibfnamefont {W.}~\bibnamefont
  {Gao}}, \bibinfo {author} {\bibfnamefont {J.}~\bibnamefont {Liu}},\ and\
  \bibinfo {author} {\bibfnamefont {F.}~\bibnamefont {Wu}},\ }\bibfield
  {title} {\bibinfo {title} {Polarization-insensitive and
  polarization-controlled dual-band third-harmonic generation in silicon
  metasurfaces driven by quasi-bound states in the continuum},\ }\href
  {https://doi.org/10.1063/5.0187055} {\bibfield  {journal} {\bibinfo
  {journal} {Applied Physics Letters}\ }\textbf {\bibinfo {volume} {124}},\
  \bibinfo {pages} {051703} (\bibinfo {year} {2024})}\BibitemShut {NoStop}%
\bibitem [{\citenamefont {Shi}\ \emph {et~al.}(2022)\citenamefont {Shi},
  \citenamefont {Gu}, \citenamefont {Zhang}, \citenamefont {Xu}, \citenamefont
  {Han}, \citenamefont {Yang}, \citenamefont {Cong},\ and\ \citenamefont
  {Zhang}}]{shi2022terahertz}%
  \BibitemOpen
  \bibfield  {author} {\bibinfo {author} {\bibfnamefont {W.}~\bibnamefont
  {Shi}}, \bibinfo {author} {\bibfnamefont {J.}~\bibnamefont {Gu}}, \bibinfo
  {author} {\bibfnamefont {X.}~\bibnamefont {Zhang}}, \bibinfo {author}
  {\bibfnamefont {Q.}~\bibnamefont {Xu}}, \bibinfo {author} {\bibfnamefont
  {J.}~\bibnamefont {Han}}, \bibinfo {author} {\bibfnamefont {Q.}~\bibnamefont
  {Yang}}, \bibinfo {author} {\bibfnamefont {L.}~\bibnamefont {Cong}},\ and\
  \bibinfo {author} {\bibfnamefont {W.}~\bibnamefont {Zhang}},\ }\bibfield
  {title} {\bibinfo {title} {Terahertz bound states in the continuum with
  incident angle robustness induced by a dual period metagrating},\ }\href@noop
  {} {\bibfield  {journal} {\bibinfo  {journal} {Photonics research}\ }\textbf
  {\bibinfo {volume} {10}},\ \bibinfo {pages} {810} (\bibinfo {year}
  {2022})}\BibitemShut {NoStop}%
\bibitem [{\citenamefont {Yan}\ \emph {et~al.}(2024)\citenamefont {Yan},
  \citenamefont {Xie}, \citenamefont {Chen},\ and\ \citenamefont
  {Wu}}]{yan2024brillouin}%
  \BibitemOpen
  \bibfield  {author} {\bibinfo {author} {\bibfnamefont {X.-F.}\ \bibnamefont
  {Yan}}, \bibinfo {author} {\bibfnamefont {Y.-J.}\ \bibnamefont {Xie}},
  \bibinfo {author} {\bibfnamefont {S.}~\bibnamefont {Chen}},\ and\ \bibinfo
  {author} {\bibfnamefont {P.-Y.}\ \bibnamefont {Wu}},\ }\bibfield  {title}
  {\bibinfo {title} {Brillouin zone folding metasurface for near perfect
  circular dichroism},\ }\href@noop {} {\bibfield  {journal} {\bibinfo
  {journal} {Advanced Optical Materials}\ }\textbf {\bibinfo {volume} {12}},\
  \bibinfo {pages} {2401027} (\bibinfo {year} {2024})}\BibitemShut {NoStop}%
\bibitem [{\citenamefont {Wang}\ \emph {et~al.}(2023)\citenamefont {Wang},
  \citenamefont {Srivastava}, \citenamefont {Tan}, \citenamefont {Wang},\ and\
  \citenamefont {Singh}}]{Wang2023}%
  \BibitemOpen
  \bibfield  {author} {\bibinfo {author} {\bibfnamefont {W.}~\bibnamefont
  {Wang}}, \bibinfo {author} {\bibfnamefont {Y.~K.}\ \bibnamefont
  {Srivastava}}, \bibinfo {author} {\bibfnamefont {T.~C.}\ \bibnamefont {Tan}},
  \bibinfo {author} {\bibfnamefont {Z.}~\bibnamefont {Wang}},\ and\ \bibinfo
  {author} {\bibfnamefont {R.}~\bibnamefont {Singh}},\ }\bibfield  {title}
  {\bibinfo {title} {Brillouin zone folding driven bound states in the
  continuum},\ }\href {https://doi.org/10.1038/s41467-023-38367-y} {\bibfield
  {journal} {\bibinfo  {journal} {Nature Communications}\ }\textbf {\bibinfo
  {volume} {14}},\ \bibinfo {pages} {2811} (\bibinfo {year}
  {2023})}\BibitemShut {NoStop}%
\bibitem [{\citenamefont {Sun}\ \emph {et~al.}(2024{\natexlab{b}})\citenamefont
  {Sun}, \citenamefont {Levy},\ and\ \citenamefont {Han}}]{sun2024exploiting}%
  \BibitemOpen
  \bibfield  {author} {\bibinfo {author} {\bibfnamefont {K.}~\bibnamefont
  {Sun}}, \bibinfo {author} {\bibfnamefont {U.}~\bibnamefont {Levy}},\ and\
  \bibinfo {author} {\bibfnamefont {Z.}~\bibnamefont {Han}},\ }\bibfield
  {title} {\bibinfo {title} {Exploiting zone-folding induced quasi-bound modes
  to achieve highly coherent thermal emissions},\ }\href@noop {} {\bibfield
  {journal} {\bibinfo  {journal} {Nano Letters}\ }\textbf {\bibinfo {volume}
  {24}},\ \bibinfo {pages} {764} (\bibinfo {year}
  {2024}{\natexlab{b}})}\BibitemShut {NoStop}%
\bibitem [{\citenamefont {He}\ \emph {et~al.}(2018)\citenamefont {He},
  \citenamefont {Guo}, \citenamefont {Feng}, \citenamefont {Xu},\ and\
  \citenamefont {Miroshnichenko}}]{He2018}%
  \BibitemOpen
  \bibfield  {author} {\bibinfo {author} {\bibfnamefont {Y.}~\bibnamefont
  {He}}, \bibinfo {author} {\bibfnamefont {G.}~\bibnamefont {Guo}}, \bibinfo
  {author} {\bibfnamefont {T.}~\bibnamefont {Feng}}, \bibinfo {author}
  {\bibfnamefont {Y.}~\bibnamefont {Xu}},\ and\ \bibinfo {author}
  {\bibfnamefont {A.~E.}\ \bibnamefont {Miroshnichenko}},\ }\bibfield  {title}
  {\bibinfo {title} {Toroidal dipole bound states in the continuum},\ }\href
  {https://doi.org/10.1103/physrevb.98.161112} {\bibfield  {journal} {\bibinfo
  {journal} {Physical Review B}\ }\textbf {\bibinfo {volume} {98}},\ \bibinfo
  {pages} {161112} (\bibinfo {year} {2018})}\BibitemShut {NoStop}%
\bibitem [{\citenamefont {Overvig}\ \emph {et~al.}(2020)\citenamefont
  {Overvig}, \citenamefont {Malek}, \citenamefont {Carter}, \citenamefont
  {Shrestha},\ and\ \citenamefont {Yu}}]{Overvig2020}%
  \BibitemOpen
  \bibfield  {author} {\bibinfo {author} {\bibfnamefont {A.~C.}\ \bibnamefont
  {Overvig}}, \bibinfo {author} {\bibfnamefont {S.~C.}\ \bibnamefont {Malek}},
  \bibinfo {author} {\bibfnamefont {M.~J.}\ \bibnamefont {Carter}}, \bibinfo
  {author} {\bibfnamefont {S.}~\bibnamefont {Shrestha}},\ and\ \bibinfo
  {author} {\bibfnamefont {N.}~\bibnamefont {Yu}},\ }\bibfield  {title}
  {\bibinfo {title} {Selection rules for quasibound states in the continuum},\
  }\href {https://doi.org/10.1103/physrevb.102.035434} {\bibfield  {journal}
  {\bibinfo  {journal} {Physical Review B}\ }\textbf {\bibinfo {volume}
  {102}},\ \bibinfo {pages} {035434} (\bibinfo {year} {2020})}\BibitemShut
  {NoStop}%
\bibitem [{\citenamefont {Lee.}\ \emph {et~al.}(2012)\citenamefont {Lee.},
  \citenamefont {Bo~Zhen}, \citenamefont {Wenjun~Qiu},\ and\ \citenamefont
  {Shapira.}}]{Lee2012}%
  \BibitemOpen
  \bibfield  {author} {\bibinfo {author} {\bibfnamefont {J.}~\bibnamefont
  {Lee.}}, \bibinfo {author} {\bibfnamefont {S.-L.~C.}\ \bibnamefont
  {Bo~Zhen}}, \bibinfo {author} {\bibfnamefont {M.~S.}\ \bibnamefont
  {Wenjun~Qiu}, \bibfnamefont {John D.~Joannopoulos}},\ and\ \bibinfo {author}
  {\bibfnamefont {O.}~\bibnamefont {Shapira.}},\ }\bibfield  {title} {\bibinfo
  {title} {Observation and differentiation of unique high-q optical resonances
  near zero wave vector in macroscopic photonic crystal slabs},\ }\href
  {https://doi.org/10.1103/physrevlett.109.067401} {\bibfield  {journal}
  {\bibinfo  {journal} {Physical Review Letters}\ }\textbf {\bibinfo {volume}
  {109}},\ \bibinfo {pages} {067401} (\bibinfo {year} {2012})}\BibitemShut
  {NoStop}%
\bibitem [{\citenamefont {Palik}(2012)}]{Palik2012}%
  \BibitemOpen
  \bibfield  {author} {\bibinfo {author} {\bibfnamefont {E.}~\bibnamefont
  {Palik}},\ }\href@noop {} {\emph {\bibinfo {title} {Handbook of Optical
  Constants of Solids}}}\ (\bibinfo  {publisher} {Academic Press},\ \bibinfo
  {year} {2012})\BibitemShut {NoStop}%
\bibitem [{\citenamefont {Sadreev}(2021)}]{sadreev2021interference}%
  \BibitemOpen
  \bibfield  {author} {\bibinfo {author} {\bibfnamefont {A.~F.}\ \bibnamefont
  {Sadreev}},\ }\bibfield  {title} {\bibinfo {title} {Interference traps waves
  in an open system: bound states in the continuum},\ }\href@noop {} {\bibfield
   {journal} {\bibinfo  {journal} {Reports on Progress in Physics}\ }\textbf
  {\bibinfo {volume} {84}},\ \bibinfo {pages} {055901} (\bibinfo {year}
  {2021})}\BibitemShut {NoStop}%
\end{thebibliography}%
\providecommand{\noopsort}[1]{}\providecommand{\singleletter}[1]{#1}%

\end{document}